\documentclass[prb,aps,twocolumn,groupaddress]{revtex4-1}
\usepackage{latexsym}
\usepackage{graphicx}
\usepackage{verbatim}
\usepackage{multirow}
\usepackage [english]{babel}
\usepackage [autostyle, english = american]{csquotes}
\MakeOuterQuote{"}
\usepackage{amsmath}
\newcommand{\beq}{\begin{eqnarray}}
\newcommand{\eeq}{\end{eqnarray}}
\usepackage{mathrsfs}
\usepackage{float}
\usepackage{caption}
\usepackage{bm}
\usepackage[usenames, dvipsnames]{color}
\usepackage{tabularx}

\usepackage{subcaption}
\usepackage{mathtools}
\usepackage{slashed}
\usepackage{physics}	
\usepackage{graphicx}   
\usepackage{epstopdf}
\usepackage{hyperref}   
\usepackage{bbold}
\usepackage{wasysym}
\usepackage{amsmath}
\usepackage{bm}

\def\2DEG{two-dimensional electron gas~}
\def\1DEG{one-dimensional electron gas~}

\def\LSCO{La$_{2-x}$Sr$_x$CuO$_4$}
\def\LBCO{La$_{2-x}$Ba$_x$CuO$_4$}

\def\C60{A$_x$C$_{60}$}

\def\BSCCO{Bi$_2$Sr$_2$CaCu$_2$O$_{8+\delta}$}

\def\HgCu3{HgCa$_2$Cu$_3$O$_{8+y}$}
\def\HgCu4{HgBa$_2$Ca$_3$Cu$_4$O$_{10+y}$}
\def\TlCu{Tl$_2$Ba$_2$CuO$_{6+\delta}$}
\def\TlCu3{Tl$_2$Ba$_2$Ca$_2$Cu$_3$O$_{10+y}$}
\def\TlCu4{Tl$_2$Ba$_2$Ca$_3$Cu$_4$O$_{12+y}$}

\def\BiCu3{Bi$_2$Sr$_2$Ca$_{2}$Cu$_3$O$_y$}

\def\8LSCO{La$_{1.88}$Sr$_{.12}$CuO$_4$}
\def\110LNSCO{La$_{1.5}$Nd$_{0.4}$Sr$_{0.1}$CuO$_{4}$}
\def\stage4LCO{La$_{2}$CuO$_{4+\delta}$}
\def\Y248{YBa$_2$Cu$_4$O$_8$}

\def\NbSe2{NbSe$_2$}
\def\TaSe2{TaSe$_2$}
\def\TiSe2{TiSe$_2$}
\def\NaCoOH2O{Na$_{0.3}$CoO$_{2y}$H$_2$O}
\def\MgB2{MgB${}_2$}

\def\FeAs122{Ca(Fe$_{1-x}$Co$_x$)$_2$As$_2$}

\begin{document}

\title{
Topology and the one-dimensional Kondo-Heisenberg model}
\author{Julian May-Mann} 
\affiliation{Department of Physics and Institute of Condensed Matter Theory, University of Illinois at Urbana-Champaign, 1110 West Green Street, Urbana, Illinois 61801-3080, USA}
\author{Ryan Levy} 
\affiliation{Department of Physics and Institute of Condensed Matter Theory, University of Illinois at Urbana-Champaign, 1110 West Green Street, Urbana, Illinois 61801-3080, USA}
\author{Rodrigo Soto-Garrido}
\affiliation{Facultad de F{\'\i}sica, Pontificia Universidad Cat\'olica de Chile, Vicu\~na Mackenna 4860, Santiago, Chile}
\author{Gil Young Cho}
\affiliation{Department of Physics, Pohang University of Science and Technology (POSTECH), Pohang 37673, Republic of Korea}
\author{Bryan K. Clark} 
\affiliation{Department of Physics and Institute of Condensed Matter Theory, University of Illinois at Urbana-Champaign,1110 West Green Street, Urbana, Illinois 61801-3080, USA}
\author{Eduardo Fradkin} 
\affiliation{Department of Physics and Institute of Condensed Matter Theory, University of Illinois at Urbana-Champaign,1110 West Green Street, Urbana, Illinois 61801-3080, USA}

\begin{abstract}
{The Kondo-Heinsberg chain is an interesting model of a strongly correlated system which has a broad superconducting state with pair-density wave (PDW) order. Some of us have recently  proposed that this PDW state  is a symmetry-protected  topological (SPT) state, and the gapped spin sector of the model supports Majorana zero modes. In this work, we reexamine this problem using a combination of numeric and analytic methods. In extensive density matrix renormalization group calculations, we find no evidence of a topological ground state degeneracy or the previously proposed Majorana zero modes in the PDW phase of this model. This result motivated us to reexamine the original arguments for the existence of the Majorana zero modes. A careful analysis of the effective continuum field theory of the model shows that the Hilbert space of the spin sector of the theory does not contain \textit{any} single Majorana fermion excitations. This analysis shows that the PDW state of the doped 1D Kondo-Heisenberg model is not an SPT with Majorana zero modes. }

\end{abstract}
\maketitle
\section{Introduction}
In recent years, evidence for nonuniform superconducting (SC) states has been found in certain high-temperature superconductors. An example of this appears to occur in the cuprate {\LBCO}  (LBCO) \cite{Li-2007,Berg-2007,Berg-2009}. At $x = 1/8$, the critical temperature $T_c$ {for the onset of the Meissner state of the uniform d-wave superconductivity is suppressed to near 4K while the resistive transition onsets at 10K. However, between 10K and 16K, where CDW and SDW orders are both present,} there is a quasi-two-dimensional SC phase, where CuO planes are superconducting but the material remains insulating along the c axis. This dynamical layer decoupling seen in LBCO {near x=1/8}, as well as in {\LSCO} (LSCO)  and LBCO in magnetic fields, can be explained if the copper oxide planes have pair-density-wave (PDW) superconducting order. In the PDW state, the superconducting order parameter oscillates in space with a given wave vector. {Further  evidence for the existence of a PDW state has been found recently in scanning tunneling microscopy experiments in the ``halo'' of superconducting vortices in {\BSCCO}. \cite{Edkins-2018} A related state was proposed quite early on by Fulde and Ferrell \cite{Fulde-1964} (FF) and independently by Larkin and Ovchinnikov \cite{Larkin-1964} (LO), who showed that it is possible to have a SC state where the Cooper pairs have nonzero center-of-mass momentum in the presence of a uniform (Zeeman) magnetic field. In contrast, the PDW state preserves time reversal symmetry, and is generated by strong electron correlations instead of a BCS like mechanism. This PDW state has also been proposed as a natural competing state of the uniform d-wave SC state in the pseudogap regime. An extensive review of the physics of PDW states and their experimental evidence is given by Agterberg and coworkers\cite{Agterberg-2019}.}

In previous work, it has been shown that a pair density wave state is supported in the doped Kondo-Heisenberg {(KH) chain, \cite{berg2010} which consists of a 1D electron gas (1DEG) coupled to a quantum Heisenberg antiferromagnetic chain by a Kondo interaction, \cite{Sikkema-1997}} and in an extended Hubbard-Heisenberg model on a two-leg ladder {at certain commensurate fillings.\cite{jaefari2012}} In the PDW phase of the KH chain, the spin degrees of freedom are gapped, {while its single charge mode  decouples and remains gapless,} and the PDW order parameter has quasi-long range order. These results have been confirmed by using powerful numerical and analytic techniques such as the density-matrix renormalization group (DMRG)  \cite{Sikkema-1997} and Abelian bosonization.  {\cite{Zachar-2001,Zachar-2001b,berg2010}  This PDW state is peculiar in that  the only allowed order parameters with quasi-long-range order are composite operators such as $\mathcal{O}_{PDW} \sim {\bm N}_h \cdot {\bm \Delta}$, where ${\bm N}_h$ is the N\'eel order parameter of the spin-1/2 Heisenberg spin chain and ${\bm \Delta}$ is the triplet superconducting order parameter of the 1DEG. All fermion bilinear observables decay exponentially with distance. Because of this feature, this PDW state cannot be described using the conventional Bogoliubov approximation, unlike the more conventional FFLO states. }

Surprisingly, in a recent publication {\cite{cho2014} three of us have put forth arguments} that in the PDW phase, the spin sector of these systems is topological and supports Majorana zero modes (MZMs). MZMs have a long history in the study of topological superconductors. In particular, MZMs are believed to exist in { vortex cores of two-dimensional} $p_x+ip_y$ superconductors,\cite{Read-2000,ivanov-2001} in quantum wires proximate to superconductors,\cite{Kitaev-2001} and in vortices on the superconducting surfaces of topological insulators.\cite{Fu-2008} In these examples, the superconductivity is encoded by use of a BCS mean field term for the fermions of the system. 

The Majorana zero modes proposed to exist in the doped Kondo-Heisenberg chain are novel in that {they originate from solitons of the spin sector of this strongly correlated system,  localized at endpoints of the chain and at junctions with conventional phases. In particular this model cannot be solved within the Bogoliubov mean field theory, in which the phase mode of the superconductor is frozen as in the case of the Kitaev wire.\cite{Kitaev-2001} If the arguments for the topological character of the PDW state of the KH chain of Ref. \onlinecite{cho2014} were correct,  the KH chain would be a natural place to test for the existence of a MZMs in a system with a dynamical massless charge mode. We should note that, after the publication of Ref.\onlinecite{cho2014}, Ruhman, Berg and Altman  have constructed a  model with protected MZMs in a (uniform) 1D superconductor with a dynamical massless phase field.\cite{Ruhman-2015} }

{In this work we reexamine the doped Kondo-Heisenberg model in detail using extensive DMRG simulations on long chains ($L=128$) with various boundary conditions.  We are able to identify the 1D PDW as was seen in Ref.~\onlinecite{berg2010} but do not find evidence of any Majorana zero modes in the PDW phase. }

{Motivated by the absence of evidence of MZMs in our numerical results, we turned to non-Abelian bosonization to reinvestigate analytically the original claims that the PDW wire is topological. In the non-Abelian bosonization approach the effective field theory of this problem consists of four dynamical Majorana fermionic fields (see also Ref.~\onlinecite{tsvelik2016b}). As anticipated in Ref.~\onlinecite{jaefari2012}, the effective field theory has two massive phases separated by a quantum phase transition in the 1+1 dimensional Ising universality class in which just one Majorana fermion becomes massless. In the massive phases all four Majorana fields are massive and are distinguished by the sign of the expectation value of the fermion bilinear of the light Majorana field. The massive phases are in the universality class of the $O(4)\simeq SU(2) \times SU(2)$ Gross-Neveu model investigated long ago by Witten \cite{Witten-1978} and by Shankar \cite{Shankar-1985}. At the critical point one Majorana fermion is massless and the remaining three Majoranas are massive and (with minor fine tuning) have an effective supersymmetry.}\cite{Witten-1978} 

{By carefully examining the full Hilbert space of the spin sector of the theory, the non-Abelian bosonization results  show explicitly that there are no states with odd-fermion parity in the physical spectrum (a necessary condition for the existence of Majorana zero modes). From this we conclude that the previously proposed Majorana zero modes do not correspond to physical operators in the doped Kondo-Heisenberg chain. Of course this result does not prove that \textit{a} PDW state cannot in principle be topological. A candidate topological PDW state is discussed qualitatively in the conclusions of this paper. Whether or not a topological PDW state is possible in a non-mean field  model with a local Hamiltonian remains an open question.} 

This paper is organized as follows. In section \ref{sec:Model} we present the model and discuss its phase diagram. In section \ref{sec:Numerics} we present our numeric analysis of the doped Kondo Heisenberg model, and the lack of evidence of the Majorana zero modes. In section \ref{sec:PrevProMZM} we present the previously proposed argument for the Majorana zero modes by using non-Abelian bosonization. In section \ref{sec:NewArg} we reexamine these claims, and show by careful analysis of the Hilbert space of the spin model that the Majorana zero modes are not physical operators. We also discuss the possibility of the doped Kondo Heisenberg model being a different {symmetry-protected topological phase (SPT)}. We conclude with a discussion of our results in section \ref{sec:Con}.
Technical parts of our analysis are presented in several appendices. In Appendix \ref{sec:NonAbel} we determine the RG equation for the Kondo Heisenberg model using non-Abelian bosonization. In Appendix \ref{sec:FermNum} we calculate the "fermion parity" of states that make up the Hilbert space of the spin sector of the theory.  In Appendix \ref{sec:ContLim} we present the continuum limit of the model using Abelian bosonization.  In Appendix \ref{sec:MajorAbel} we use Abelian bosonization to show that the proposed Majorana zero modes are not physical operators. In Appendix \ref{sec:OrderParam} we discuss the order parameters that differentiate the trivial and PDW phases of the model.

\section{Model and PDW states}
\label{sec:Model}

In previous work, it has been shown that a PDW phase exists in the doped 1D Kondo-Heisenberg ladder\cite{berg2010}. The Kondo-Heisenberg ladder consists of a 1D electron gas (1DEG) coupled to a Heisenberg spin-$1/2$ chain via Kondo couplings. The Hamiltonian for this system is
\begin{equation}
\begin{split}
\mathcal{H} =& \mathcal{H}_{e}+\mathcal{H}_{H}+\mathcal{H}_{K}\\
\mathcal{H}_{e} =& -t\sum_{j,\sigma} c^\dagger_{j,\sigma}c_{j+1,\sigma} +h.c.\\ \phantom{=} & + U\sum_{j}{n}_{\uparrow} {n}_{\downarrow} - \mu\sum_{j,\sigma}n_{j,\sigma}\\
\mathcal{H}_{H} =& J_H \sum_j \bm{S}_{j,h} \cdot  \bm{S}_{j+1,h}+ J'_H \sum_j \bm{S}_{j,h} \cdot  \bm{S}_{j+2,h}\\
\mathcal{H}_{K} =& J_K\sum_j \bm{S}_{j,h} \cdot  \bm{S}_{j,e},
\label{eq:LatHam}
\end{split}
\end{equation}
where $c^\dagger_{j,\sigma}$ are the electron creation operators, $\bm{S}_{j,h}$ are the Heisenberg spin operators, $
\bm{S}_{j,e} = \frac{1}{2}c^\dagger_{j,\sigma} \bm{\tau}_{\sigma,\sigma'} c_{j,\sigma'}$ are the electron spin operators, and $\bm{\tau}$ are the Pauli matrices. We have included additional Hubbard $U$ interactions for the 1DEG and a next nearest neighbor spin coupling $J'_H$ in the Heisenberg chain. We will consider the case where the 1DEG electrons have been doped away from half filling. This model also arises naturally in two leg Hubbard ladders, where the bonding band is at half filling\cite{jaefari2012}. In this case, the Umklapp process gaps-out the charge degrees of freedom in the bonding band, and the Kondo and Heisenberg couplings for the spin degrees of freedom are generated perturbatively. 

In terms of the spin and charge currents of the system, the continuum limit of Eq.~\ref{eq:LatHam} is given by
\begin{equation}
\begin{split}
\mathcal{H} =& \mathcal{H}_{c}+\mathcal{H}_{s}\\
\mathcal{H}_{c} =& \frac{\pi v_c}{2}[ J_{e,R} J_{e,R}+ J_{e,L} J_{e,L} ] +g_c J_{e,R}J_{e,L}\\
\mathcal{H}_s =& \frac{2\pi v_{s,e}}{3} \bm{J}_{e,R} \bm{J}_{e,R} + \frac{2\pi v_{s,h}}{3} \bm{J}_{h,R} \bm{J}_{h,R} + (R\leftrightarrow L)
\\ &  -g_{s1} [\bm{J}_{e,R} \bm{J}_{e,L} + \bm{J}_{h,R} \bm{J}_{h,L} ] \\&-g_{s2} [\bm{J}_{e,R} \bm{J}_{h,L} + \bm{J}_{h,R} \bm{J}_{e,L} ],
\label{eq:BosonHam0}
\end{split}
\end{equation}
where $J_{e}$ are the electron $U(1)$ charge currents, and $\bm{J}_{e/h}$ are the {1DEG and Heisenberg chain} $SU(2)$ spin currents, respectively. The Abelian bosonization of this model, and weak coupling analysis is discussed in Appendix \ref{sec:ContLim}. The phase diagram for this system has been previously determined using Abelian bosonization\cite{jaefari2012}, and are rederived here using non-Abelian bosonization\cite{Witten-1984} in Appendix \ref{sec:NonAbel}. Eq. \ref{eq:BosonHam0} has three fixed points corresponding to $(g_{s1},g_{s2}) = (0,0)$, $(-\infty,0)$, $(0,-\infty)$. When $(g_{s1},g_{s2}) = (0,0)$ the system is a Luttinger liquid with 1 charge degree of freedom and 2 spin degrees of freedom {(a C1S2 Luttinger liquid in the terminology of Ref.~\onlinecite{Balents-1996}.)}

At the $(g_{s1},g_{s2}) = (0,-\infty)$ fixed point, the system is in a PDW phase, since the PDW order parameter 
\beq
\nonumber O_{PDW} = && \bm{\Delta}\cdot \bm{N}_h
\label{eq:PDWOrder}
\eeq
has quasi-long range order. Here, $\bm{\Delta}$ is the triplet superconductivity order parameter {of the 1DEG}, and $\bm{N}_h$ is the staggered (N\'eel) component of the magnetization of the Heisenberg spins. In addition, the singlet superconducting order parameter decays exponentially fast. In the PDW phase the charge sector remain gapless, while the spin sector acquire a gap, and the magnetization vanishes in the ground state, $S^z \equiv  \sum_j S^z_{j,e} + S^z_{h,e}  = 0$. At the $(g_{s1},g_{s2}) = (-\infty,0)$ fixed point, the system is in a conventional SC phase, since the the singlet SC order parameter has quasi-long range order, while the PDW order parameter decays exponentially fast. In this conventional SC phase, the charge sector is also free and the spin sector is gapped with vanishing magnetization. The line $g_{s1} = g_{s2} < 0$ marks a quantum phase transition between the PDW and trivial SC phases that is in the Ising universality class and can be described in terms of a free Majorana fermion.

In previous works, it has been argued that in the PDW phase, the gapped spin sector of the model is topological and hosts Majorana zero edge modes\cite{cho2014}. The `fermion parity' associated with a pair of these Majorana zero modes corresponds to the relative spin parity of the lattice model 
\beq
(-1)^{ Q^z}, \phantom{=} Q^z \equiv \sum_j S^z_{j,e}-S^z_{j,h}.
\label{eq:spinPar1}
\eeq

\section{Numerics}\label{sec:Numerics}

In this section, we will use Density Matrix Renormalization Group\cite{schollwock2011} (DMRG) to search for evidence of the Majorana edge modes eventually concluding that the numerics do not support the existence of Majorana edge modes in the PDW phase. 

We start by considering the Hamiltonian in Eq. \ref{eq:LatHam} on a finite ladder with $L=32-128$ rungs, $n=0.875$ filling, and open boundary conditions.  We primarily consider the parameters $t=1, J_H=J_K=2, J_H^\prime=U=0$ which correspond to those used in Ref.~\onlinecite{berg2010}. We obtain a ground state and first excited state, keeping up to $m=7200$ states with truncation errors $< 10^{-8}$ and $\braket{\psi_1}{\psi_0} < 10^{-7}$. 
\begin{figure}[hbt] 
     \centering
     \includegraphics[width=\columnwidth]{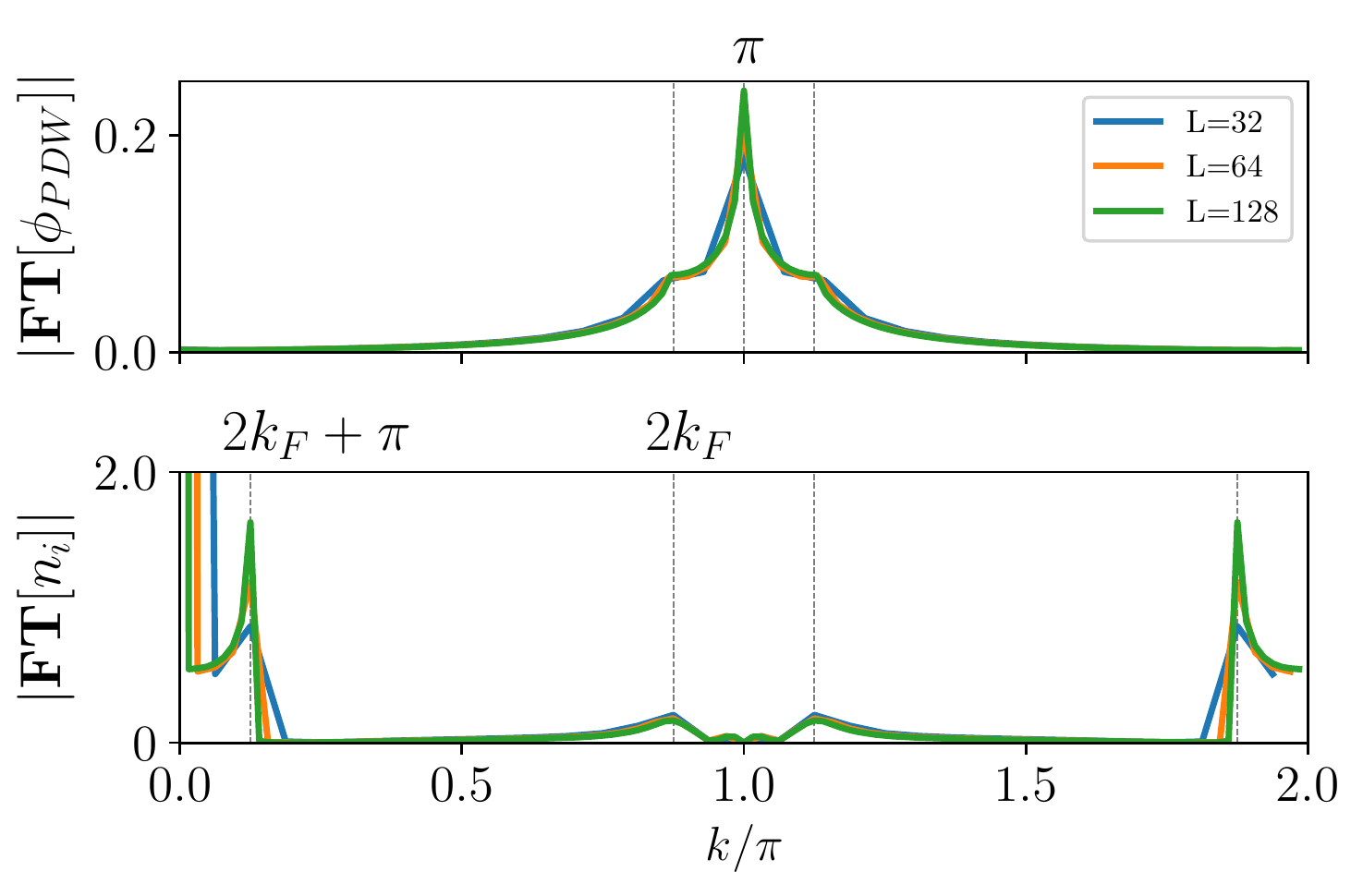}
     \caption{Magnitude of the Fourier transform of the PDW correlation function (\textit{top}) and charge density (\textit{bottom}) of the ground state. $\phi_{PDW}$ was averaged over all possible $|i-j|$.}
     \label{fig:pdw_features}
\end{figure}

We first validate that we get the PDW in the ground state.  We measure the order parameters,
\begin{align}
    \phi^\dagger_{B,i} =& \frac{1}{2} \left(c^\dagger_{i\uparrow}c^\dagger_{i+1\downarrow}-c^\dagger_{i\downarrow}c^\dagger_{i+1\uparrow}\right)\\
   \phi_{PDW}=&\left\langle (-1)^{|i-j|} \phi^\dagger_{B}\phi_{B} (|i-j|) \right\rangle
\end{align}
In Fig \ref{fig:pdw_features}, we see the salient features of the PDW quasi-long range order - the oscillation of the $\phi_B$ bond singlet order and an accompanying charge density wave. Thus with open boundary conditions we're able to obtain the proper phase. 

There are a number of ways to establish the existence of Majorana zero edge modes (MZEMs).  To begin with, such a system will have degenerate energy eigenstates in the thermodynamic limit.  The two degenerate eigenstates will be topological and naively should have different parity values (Eq. \ref{eq:spinPar1}) as well as identical local reduced density matrices in the bulk. The edges of the two eigenstates would naturally show edge modes that should be visible in the spin-order near the location of the Majoranas.  
An additional signature of these edge modes is the existence of degeneracy in the entanglement spectrum \cite{Pollmann-2010} of the ground state. While these attributes typically hold only for gapped systems, we presume that the gapless charge mode would sufficiently decouple and not affect these properties. 

\begin{figure}[hbt]
\noindent \centering{}\includegraphics[width=1.0\columnwidth]{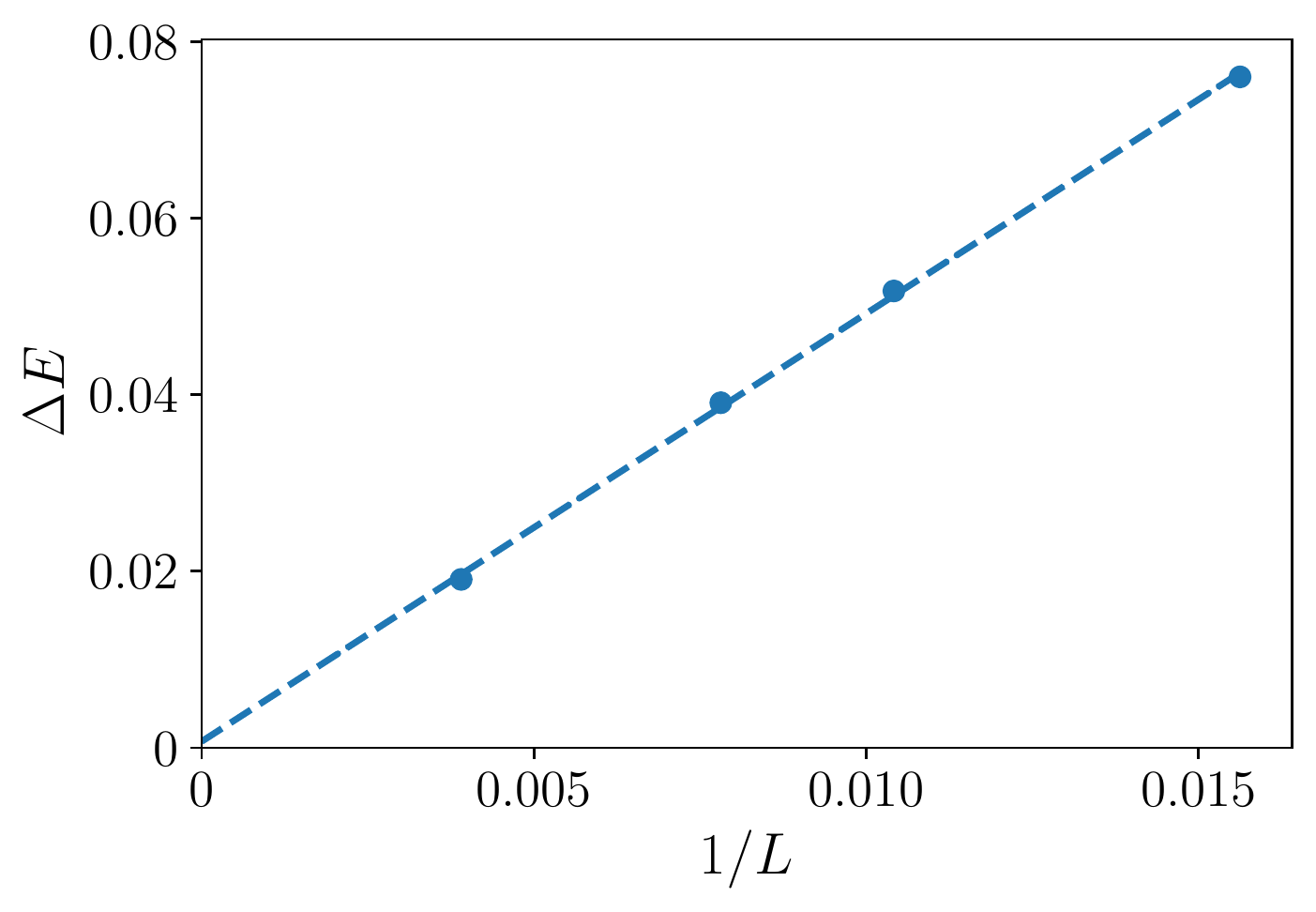}
\caption{Finite size scaling of the energy gap. For each point we  variance extrapolate near the end of DMRG optimization (see Figs.~\ref{fig:varience_E0} and~\ref{fig:varience_E1}). The linear fit gives a thermodynamic gap of $\Delta E(L=\infty)\sim 0.0007 \approx 0$.}
\label{Fig:TowerOfStates} 
\end{figure}

We begin by searching for the two degenerate states;  Ref. \onlinecite{berg2010} finds a spin gap to other $S_z$ sectors and therefore we would anticipate that the degenerate state should be in the $S^z=0$ sector although everything in this sector has parity 1. 
States which are degenerate in the thermodynamic limit, will split in energy in any finite system.  This energy splitting should (for large enough systems) decay exponentially with system size.  Therefore to search for the topological pair of states we calculate the lowest two $S^z=0$ eigenstates and look at the energy as a function of system size out to $L=128$.  Instead of an exponentially decaying gap, we  find a gap which is linear in $1/L$ extrapolating to zero in Fig.~\ref{Fig:TowerOfStates};  this is exactly what is expected for the tower of states coming from a gapless charge density wave. 

\begin{figure}[hbt]
    \centering
     \includegraphics[width=\columnwidth]{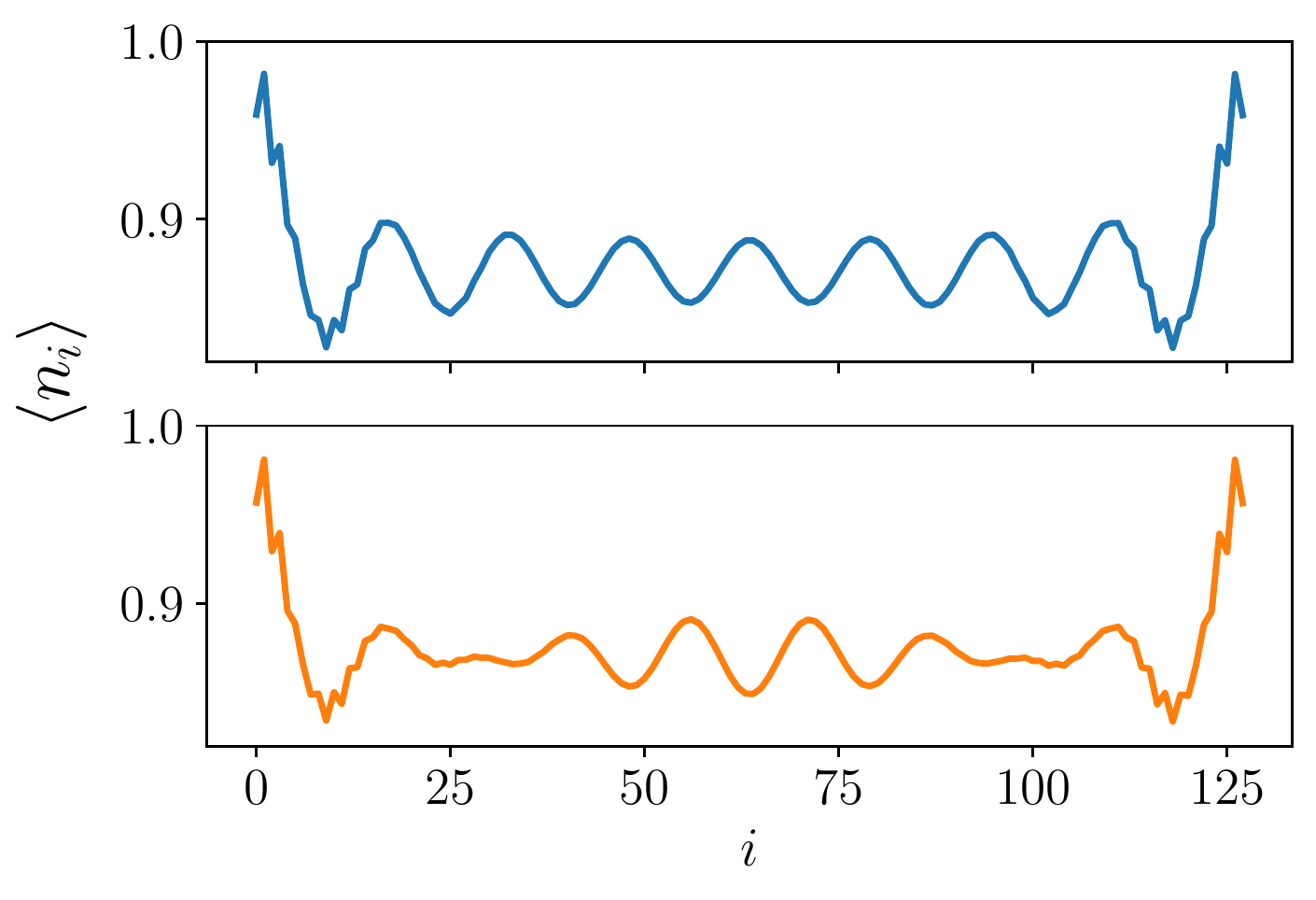}
     \caption{ Charge density in the ground state  (\textit{top}) and first excited state (\textit{bottom}) for $L=128$ and $n=0.875$.}
     \label{fig:chargeModes}
\end{figure}

In spite of this fact, we can compare these two eigenstates.  We find that the charge density of the two  eigenstates look very different (see Fig.~\ref{fig:chargeModes}) ruling out they could be topological pairs.   


\begin{figure}[hbt]
    \centering
     \includegraphics[width=\columnwidth]{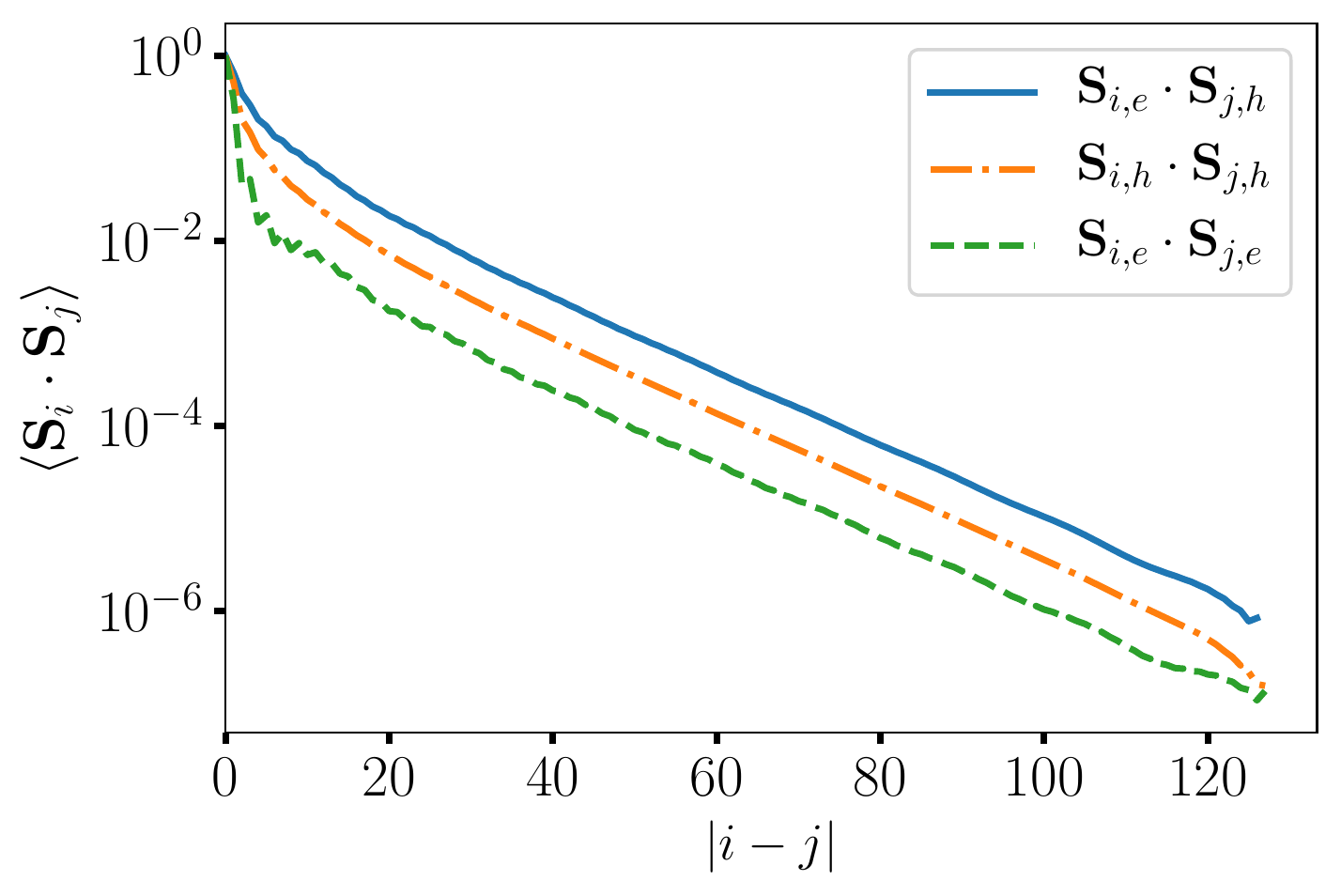}
     \caption{Normalized spin-spin correlation between parts of the ladder for $L=128$ ground state. $\langle \mathbf{S}_i \cdot \mathbf{S}_j \rangle$ was averaged over all possible $|i-j|$. }
     \label{fig:SpinSpin}
\end{figure}

\begin{figure}[hbt]
\centering{}
    \begin{subfigure}[b]{\linewidth}
    \noindent\centering{}\includegraphics[width=1.0\columnwidth]{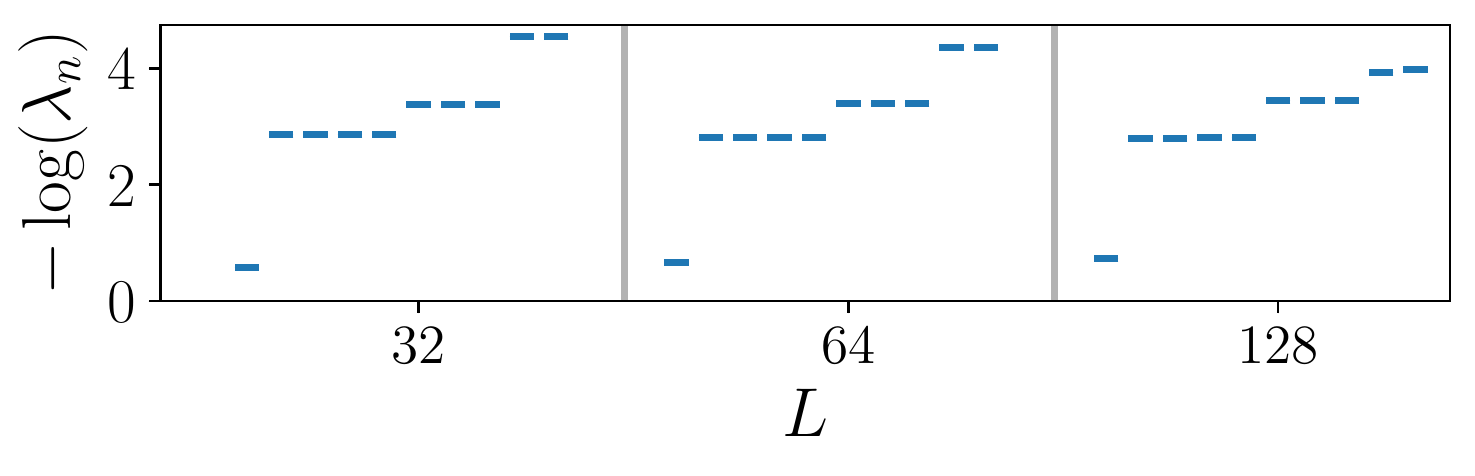}
    \caption{Open Boundary }
    \label{Fig:EntropyFinite}
    \end{subfigure}
~
    \begin{subfigure}[b]{\linewidth}
    \noindent\centering{}\includegraphics[width=0.4\columnwidth]{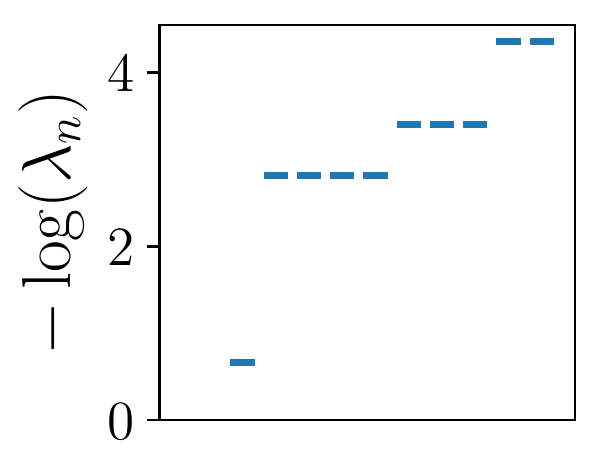}
    \caption{Insulator-PDW-Insulator Sandwich}
    \label{Fig:EntropySandwich}
    \end{subfigure}
\caption{Entanglement eigenvalue spectrum between the left and right half of the system for two boundary conditions: open (\textit{top}) and sandwich (\textit{bottom}).  }
\label{Fig:EntropyScaling}
\end{figure}

It is clear then that we don't find the topological eigenstates out to this system size.  We can also just look at the properties of only the ground state in the hope that the topological state is still too high in energy. Similar to a Haldane phase, one might find spin features localized near the edge/interface or spin-spin correlations peaked near the edge. In the ground state, the expectation values $\langle S^z_{j,e} \rangle$ and $\langle S^z_{j,h} \rangle$ are always very small (less than $10^{-8}$ in magnitude) indicating an absence of any edge-mode. In addition, there are no significant edge-edge spin-spin correlations, as seen in Fig~\ref{fig:SpinSpin}.  We also can consider the entanglement spectrum (see Fig.~\ref{Fig:EntropyScaling}) and find that the lowest entanglement eigenvalues are non-degenerate, unlike what would be anticipated for a topological system.

\begin{figure}[hbt]
    \centering
     \includegraphics[width=\columnwidth]{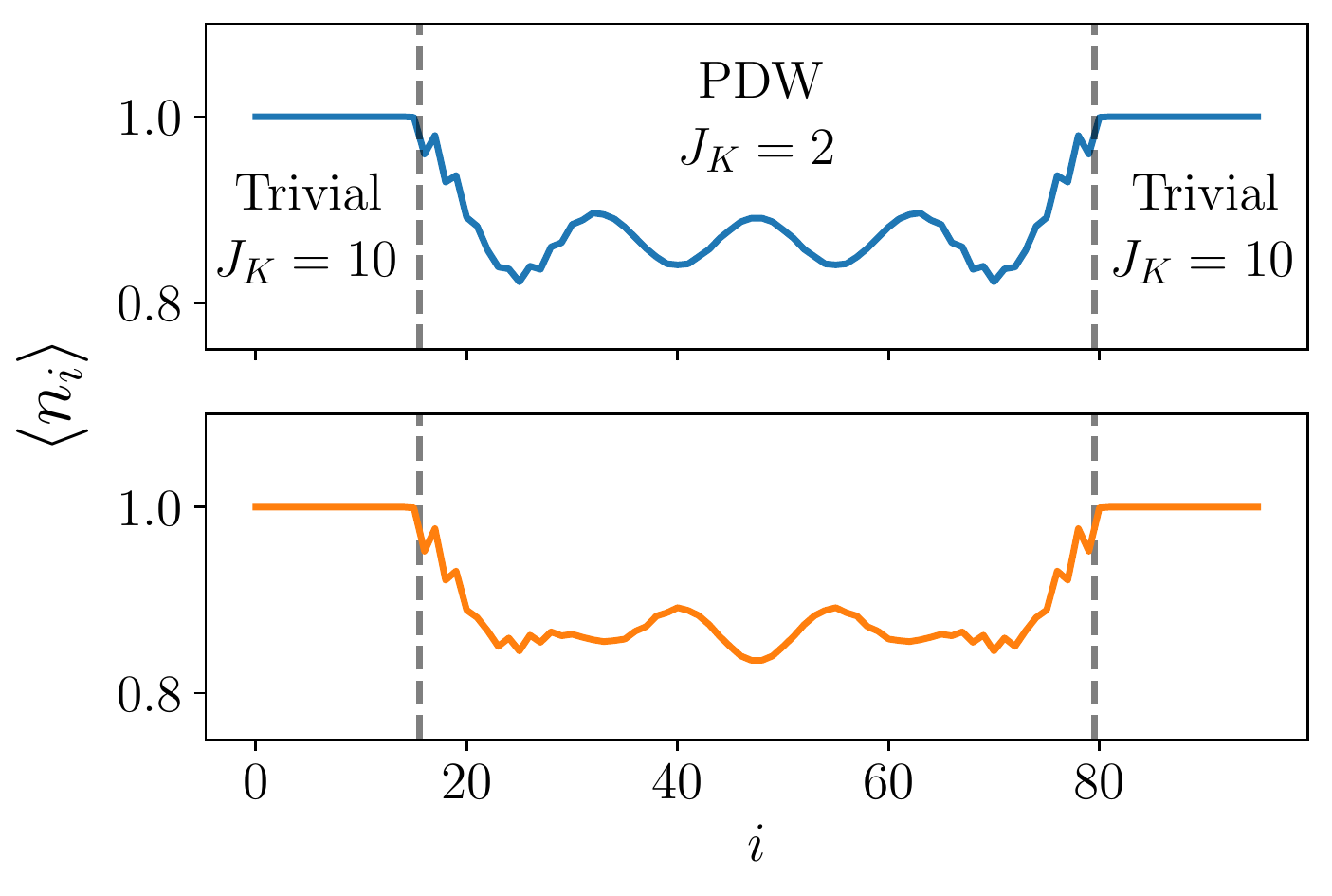}
     \caption{ Charge density  in the ground state  (\textit{top}) and excited state (\textit{bottom}) of the sandwich. }
     \label{Fig:chargeModesSandwich}
\end{figure}

As a final search, we consider sandwiches, where we vary the value of $J_K$ in different sections of the ladder. Sandwiches have been found to be helpful in identifying non-topological zero modes in Ref.~\onlinecite{robinson2019}.  Here we considered a sandwich with PDW in the bulk ($J_K=2$) and an insulator phase ($J_K=10$) on the left and right 16 rungs (see Fig.~\ref{Fig:chargeModesSandwich}). We maintain doping in the 1DEG such that the left and right insulators are half filled and the bulk maintains $\langle n \rangle \approx 0.875$. We do find PDW in the bulk as expected and explore for the presence of a Majorana mode in the interface of our sandwich.  We again consider the ground and excited state. The gap is small ($\approx 0.0395t$, which we choose not to extrapolate for computational considerations) nearly the same as the open boundary condition gap ($\approx 0.0392t$). The charge-density, shown in Fig.~\ref{Fig:chargeModesSandwich}, looks very different in the bulk suggesting the states aren't topological. We also consider the entanglement entropy in Fig.~\ref{Fig:EntropySandwich}, which has a nearly identical entanglement spectrum to the open boundary system. 

All of the evidence presented does not provide any numerical evidence of MZMs. Despite clearly finding a PDW for both open and sandwich boundary conditions, neither the ground state nor excited state of those systems show topological behavior. 

\section{Previously Proposed Majorana Zero Modes}
\label{sec:PrevProMZM}

Due to the lack of numeric evidence of MZMs, we will reexamine the arguments that the PDW wire is topological. The original argument was made using Abelian bosonization and subsequent refermionization.\cite{cho2014} Here, we shall rederive these results using non-Abelian bosonization, since it is better suited to study the non-Abelian $SU(2)$ currents of the spin sector. Similar calculations have been previously done by Tsvelik.\cite{tsvelik2016,tsvelik2016b}


There are three currents to study when considering the Kondo-Heisenberg model. A $U(1)_2$ current describing the charge degrees of freedom of the 1DEG, a $SU(2)_1$ current describing the spin degrees of freedom of the 1DEG, and a second $SU(2)_1$ current describing the Heisenberg spins, leading to a total current structure of $U(1)_2\times SU(2)_1\times SU(2)_1$, as shown in Eq. \ref{eq:BosonHam0}. {Since in the low energy limit the  charge and spin sectors decouple (spin-charge separation) , we will only focus on the spin sector,} which corresponds to a $SU(2)_1\times SU(2)_1$ Wess-Zumino-Witten (WZW) model.\cite{Witten-1984} It will also be useful to define the following currents,
\beq
\nonumber \bm{J}_{\pm,R} =  \bm{J}_{e,R} \pm \bm{J}_{h,R} \\
\bm{J}_{\pm,L} =  \bm{J}_{e,L} \pm \bm{J}_{h,L}.
\label{eq:NABCurr1}
\eeq
Here the $\bm{J}_{+}$ fields describe the $SU(2)_2$ currents, and $\bm{J}_{-}$ describe the remaining $SU(2)_1 \times SU(2)_1/SU(2)_2$ currents. In terms of these fields, the spin Hamiltonian $H_\text{s}$ becomes ({after setting the velocities of the spin modes to be equal to each other,} $v_{s,t} = v_{s,b} = v_s$)
\beq
\nonumber \mathcal{H}_{\text{s}} = &&\frac{2\pi v_s}{6} [\bm{J}_{+,R} \bm{J}_{+,R} +\bm{J}_{-,R} \bm{J}_{-,R}]\\
&-& g_{+} \bm{J}_{+,R} \bm{J}_{+,L}  - g_{-} \bm{J}_{-,R} \bm{J}_{-,L}.
\label{eq:NABSpin2}
\eeq
where $g_{\pm} = (g_{s1} \pm g_{s2})/2$. 

Using the RG equations for Eq. \ref{eq:NABSpin2} (see Appendix \ref{sec:NonAbel}), we can identify the four fixed points $(g_+,g_-) = (0,0)$, $(-\infty,\infty)$, $(-\infty,-\infty)$, $(-\infty,0)$. The $(g_+,g_-) = (0,0)$ fixed point corresponds to the $C1S2$ Luttinger state, the $ (g_+,g_-) =(-\infty,\infty)$ fixed point corresponds to the PDW phase and the $ (g_+,g_-) = (-\infty,-\infty)$ fixed point corresponds to the trivial SC phase. The $(g_+,g_-) = (-\infty,0)$ fixed point marks the Ising transition between the PDW and trivial SC phase.

To probe the existence of Majorana zero modes, we note that the two $SU(2)$ currents of the spin sector are equivalent to a single $SO(4)$ current since $SU(2)\times SU(2) \cong SO(4)$. The $SO(4)$ current algebra can naturally be expressed in terms of 4 Majorana fermions. With this in mind, let us now introduce the Majorana fermions $\eta_{0,R(L)}$  and $\eta_{a,R(L)}$, where $a = 1,2,3$. Using them, we can construct the left and right moving currents $\bm{J}_{\pm, R(L)}$ as 
\beq
\nonumber J^a_{+,R} &=& \frac{i}{2}\epsilon^{abc}\eta_{b,R}\eta_{c,R}\\
J^a_{-,R} &=& i \eta_{0,R}\eta_{a,R}. 
\label{eq:MajCurrent}
\eeq

In terms of the Majorana fermions, the spin Hamiltonian becomes 
\begin{equation}
\begin{split}
\mathcal{H}_{\text{s}} =& \frac{iv_s}{2} (\eta_{0,L} \partial_x \eta_{0,L}-\eta_{0,R} \partial_x \eta_{0,R})\\ \phantom{=}&+ \frac{iv_s}{2} \sum_{a} (\eta_{a,L} \partial_x \eta_{a,L}-\eta_{a,R} \partial_x \eta_{a,R})\\ \phantom{=}& - g_+\sum_{a>b} (\eta_{a,R} \eta_{a,L})(\eta_{b,R}\eta_{b,L})\\ \phantom{=}& - g_{-} (\eta_{0,R}\eta_{0,L})\sum_{a}(\eta_{a,R}\eta_{a,L}).
\label{eq:4MajHam1}
\end{split}
\end{equation}
{which, upon setting $g_+=g_-$, is the Hamiltonian of the $O(4)$ Gross-Neveu model. Notice that in the full problem of Eq.\eqref{eq:4MajHam1}, the ``light'' Majorana field $\eta_0$ becomes massless at $g_-=0$ and decouples from the rest.} Due to the single free Majorana fermion, $g_-=0$ marks an Ising critical point. We discuss the associated $\mathbb{Z}_2$ symmetry breaking that occurs at the phase transition in appendix \ref{sec:OrderParam}. 

In addition, this system also has a conserved fermion parity, which can be expressed as
\begin{equation}
\begin{split}
(-1)^{N_f} = \exp (i \pi \int dx [i\eta_{0,R}\eta_{3,R}+i\eta_{1,R}\eta_{2,R}+ (R\leftrightarrow L)])
\end{split}
\label{eq:fermParNA}
\end{equation}
In terms of the lattice degrees of freedom, $(-1)^{N_f} = (-1)^{\sum_j 2 S^z_{j,e}}$, which reduces to Eq. \ref{eq:spinPar1} in the ground state, where $\sum_j [S^z_{j,e}+S^z_{j,h}] = 0$.

When $g_+$ is large, we expect that $i\eta_{a,R}\eta_{a,L}$ will gain an expectation value $\langle i\eta_{a,R}\eta_{a,L}\rangle = \Delta$. With this substitution, Eq. \ref{eq:4MajHam1} becomes, 
\beq
\mathcal{H}_{\text{s}} &=& \frac{iv_s}{2} (\eta_{0,L} \partial_x \eta_{0,L}-\eta_{0,R} \partial_x \eta_{0,R})\\\nonumber &\phantom{=}& + ig_{-} \Delta(\eta_{0,R}\eta_{0,L}).
\label{eq:4MajHam2}
\eeq
Between the PDW phase ($g_{-}>0$) and the trivial SC phase ($g_{-}<0$), the mass term for $\eta_{0}$ changes sign, and one would expect for there to be a localized Majorana zero mode {at the open ends of the system.}

\section{New Arguments}
\label{sec:NewArg}

As we have shown in the previous section, the spin degrees of freedom of the doped Kondo-Heisenberg model can be expressed in terms of {four Majorana fermionic fields}. Based on this, {it is reasonable to conjecture, as was done in Ref.~\onlinecite{cho2014}, that there may be} Majorana zero modes at interfaces between the PDW and trivial SC phases. However, as we shall argue below, these is not the case here, and the doped Kondo-Heisenberg model in the PDW phase does not host Majorana zero modes. 

Let us first review several well known features of SPTs. First SPTs are short range entangled {gapped} states of matter that cannot be smoothly deformed into a trivial state   while preserving both symmetries and the bulk gap of the system. Second, at the interface between an SPT and a trivial state, there are localized zero energy degrees of freedom. This leads to a robust ground state degeneracy for a system with symmetry preserving boundaries. 

In the case of the fermionized spin sector of the doped Kondo-Heisenberg model, the localized zero energy modes are Majorana zero modes, and the ground state degeneracy corresponds to the two fermion parity sectors. Acting on a ground state with a Majorana zero mode changes the {fermionic} parity of the ground state from $\pm 1$ to $\mp 1$. Importantly, having two distinct fermion parity sectors is a necessary condition for the existence of Majorana zero modes. In reverse, if all states in a given theory have the same fermion parity, then a single Majorana zero mode is not a physical operator. 

{The underlying question we are asking is if the Hilbert space of the spin sector of the original model, Eq. \ref{eq:BosonHam0}, is the same as that of the fermionized model, Eq. \ref{eq:4MajHam1}. Clearly, the Hilbert space of the fermionized model will consists of states with both even and odd fermion parity. In the following, we will discuss whether or not both of these fermion parity sectors exist in the Hilbert space of the original spin model. We find that all states in the Hilbert space of the spin model have even fermion parity. This means that  the Hilbert space of the fermionic theory of Eq. \ref{eq:4MajHam1} is larger than that of the spin sector of the Kondo-Heisenberg model. In particular,  there are extra, unphysical, states with odd fermion parity, that do not correspond to any state in the physical Hilbert space of the spin model. A similar situation is well known to happen in the quantum Ising chain which is described by the parity even sector of the fermionized version of the model.}

To show this, it will be useful to define the system on a ring of length $L$. We are only interested in the topological features of the spin sector of the theory (Eq. \ref{eq:NABSpin2}). In order to have a pair of Majorana zero, we will put half of the ring in the PDW phase ($g_- > 0$ for $0<x<L/2)$ and the other half in the trivial SC phase ($g_- < 0$ for $L/2<x<L$). From our earlier analysis, we expect that there will be two Majorana zero modes located at $0$ and $L/2$. Since there are two Majorana zero modes in this system, we expect that there will be two degenerate ground states, one with fermion parity $+1$ and one with fermion parity $-1$. 

With this system in mind, we now ask if the fermion parity odd states exists in the Hilbert space of the model described above. In order to probe this Hilbert space, it will actually be sufficient to just probe the Hilbert space of the unperturbed model ($g_- = g_+ = 0$), which is simply the $SU(2)_1 \times SU(2)_1$ WZW model. If all states in the Hilbert space of the unperturbed model have the same fermion parity, then all states in the Hilbert space of the perturbed model will also have the same fermion parity. This is because turning on a perturbation cannot add new states to the Hilbert space. 

It is well known that the Hilbert space of a $1+1$D CFT can be organized into Verma modules that are built off of a highest weight state \cite{Difrancesco1997}. These highest weight states are created by acting on the vacuum of the theory with a primary field. In appendix \ref{sec:FermNum}, we explicitly calculated the fermion parity (Eq. \ref{eq:fermParNA}) of all states in all Verma modules of the $SU(2)_1 \times SU(2)_1$ WZW CFT. We find that they all have even fermion parity, and, as a result, all states in the perturbed model must also have even fermion parity. Individual Majorana zero mode operators are therefore not physical operators since acting on an even fermion parity state with the Majorana zero mode operator leads to an odd fermion parity state, the latter of which we know does not exist in the $SU(2)_1 \times SU(2)_1$ theory. Products of an even number of Majorana operators are physical, as can be seen from examining the $SU(2)$ currents of the model. 

From this analysis, we can conclude that switching from the spin currents (Eq. \ref{eq:NABSpin2}) to the fermion representation (Eq. \ref{eq:4MajHam1})  introduces new states into the Hilbert space of the system. In particular, the fermion parity-odd states are part of the unphysical fermionic Hilbert space, but not of the physical spin Hilbert space. So, in order move from the expanded fermionic Hilbert space to the physical spin Hilbert space, the fermionic Hilbert space must be projected onto the fermion parity even states (known in string theory as a GSO projection \cite{Polchinski-1998}). We present a similar argument using Abelian bosonization in Appendix \ref{sec:MajorAbel}.

We can also consider the possibility that the spin sector of the doped Kondo-Heisenberg model is another SPT protected by some other symmetry. The only other symmetry in the model is the total spin $SU(2) \cong SO(3)$ symmetry of the model. From cohomology classifications, it is known that there is one non-trivial SPT in 1d protected by the $SO(3)$ symmetry--the Haldane phase of the spin 1-chain. It is known that in the Haldane phase, the edge modes carry spin-1/2. In the Majorana representation only the fermions $\eta_{a}$ ($a = 1,2,3$) carry spin. It is clear that there are no zero modes for $\eta_{a}$ in Eq. \ref{eq:4MajHam1} at a boundary between the PDW and trivial SC phases, since $g_+<0$ for both phases. This indicates that the spin sector of the model is not in the Haldane phase. 

In addition, it is known that the $SU(2)_1 \times SU(2)_1$ WZW model enters the Haldane phase when the following interaction is added\cite{Affleck-1987,shelton1996,allen1997,lecheminant2002}:
\beq
H_{\text{int}} =  \frac{\lambda}{2\pi} \sum_a \text{tr}(g_e \tau^a)\text{tr}(g_h \tau^a),
\label{eq:HalCoup}
\eeq
where $g_{e/h}$ are the WZW $g$ fields of the {1DEG} and Heisenberg spins respectively (see Appendix \ref{sec:NonAbel}), and $\lambda$ is negative. In terms of the fermionic representation, this interaction introduces a negative mass terms for $\eta_{a}$ and, by extension, three Majorana zero modes at the boundaries of the system. These zero modes carry spin as expected in the Haldane phase. As shown in Appendix \ref{sec:NonAbel}, the interaction in Eq. \ref{eq:HalCoup} is not present in the doped Kondo Heisenberg model. Because of this, we can conclude that the doped Kondo Heisenberg model is not in the Haldane phase, and thereby is not an SPT.

\section{Conclusion}\label{sec:Con}
In this work, we have established using both numeric and analytic methods that the doped Kondo Heisenberg model does not host Majorana zero modes. Furthermore, it appears that the spin sector of the model is also not an SPT protected by the $SO(3)$ symmetry of the model. Based on this, we believe that the  doped Kondo Heisenberg model is not an SPT of any kind. Our analysis does not rule out possible SPTs that exist beyond the cohomology classifications, however, there is no evidence for this, and we believe that this situation is extremely unlikely. 

While our result do show that the PDW state of the Kondo Heisenberg model in 1D is not topological, it does not rule out a topological PDW state in principle. Indeed, it is easy to imagine a 1D toy model with properly chosen PDW mean field term that would have Majorana zero modes analogous to the Kitaev chain. {Since in dimensions $d>1$ PDW states generally have Fermi surfaces of Bogoliubov quasiparticles, in 1D one would expect that a PDW should have Majorana ``zero-modes'' along the length of the state. One such example is a paired $p$-wave state whose order parameter changes periodically its sign, i.e. a PDW relative of the uniform $p$-wave state. This state can be viewed as a sequence of regions with local uniform $p$ wave order with a periodic arrangement of domain walls where the sign changes occur.  Then, a Jackiw-Rebbi type argument \cite{Jackiw-1976}  implies the existence of (Majorana) zero modes at the location of each domain wall. A related topological two-dimensional state was recently studied by Santos and collaborators.\cite{santos2019pair} Actually, such a $p$-wave PDW is equivalent to a theory of massless Majorana fermions and is at a critical point. Subsequent breaking of inversion symmetry (by a uniform $p$ wave component) leads to a gapped topological state. It would be interesting to construct a 1D Hamiltonian with a state of this type (without resorting to a proximity effect mechanism). }

{Moving on to two dimensions, it is not difficult to imagine a weak-coupling 2D topological FFLO-type state. For example, if two spin-filtered Fermi surfaces exist away from the gamma point, as like the Fermi surface of doped transition metal dichalcogenides, and if there is an intra-valley triplet pairing channel, then its  natural ground state should be an intra-valley p-wave SC. Such a state is topological. The resulting topological content will be $\text{Ising} \times\bar{\text{Ising}}$. Note that this state can melt into the two distinct states, an isotropic 4e superconducting state and a CDW state without superconductivity. The topological nature of these states may be interesting to study in future work. On the other hand, since non-mean-field 2D models of PDW systems remain elusive, it is an open question whether  topological PDW states may exist in higher dimensions.} An effective field theory approach using a non-linear sigma model may be a promising way to probe this question in future work.

\section{Acknowledgments}

We thank H. Goldman and R. Sohal for useful conversations. We thank Jahen Claese, Xiongjie Yu, and Han-Yi Chou for preliminary work on simulating PDW phases.  JMM is supported by the National Science Foundation Graduate Research Fellowship Program under Grant No. DGE – 1746047. DMRG Calculations used the ITensor Library\cite{itensor}. This project is part of the Blue Waters sustained petascale computing project, which is supported by the National Science Foundation (awards OCI-0725070 and ACI-1238993) and the State of Illinois. Blue Waters is a joint effort of the University of Illinois at Urbana-Champaign and its National Center for Supercomputing Applications.  
This work also made use of the Illinois Campus Cluster, a computing resource that is operated by the Illinois Campus Cluster Program (ICCP) in conjunction with the National Center for Supercomputing Applications (NCSA) and which is supported by funds from the University of Illinois at Urbana-Champaign. This work was supported in part by the National Science Foundation grant No. DMR-1725401 at the University of Illinois (EF), and Fondecyt (Chile) Grant No. 1200399 (R.S.-G.)

\appendix 

\renewcommand{\thefigure}{S\arabic{figure}}
\setcounter{figure}{0}
\begin{figure}
    \centering
    \includegraphics[width=0.8\columnwidth]{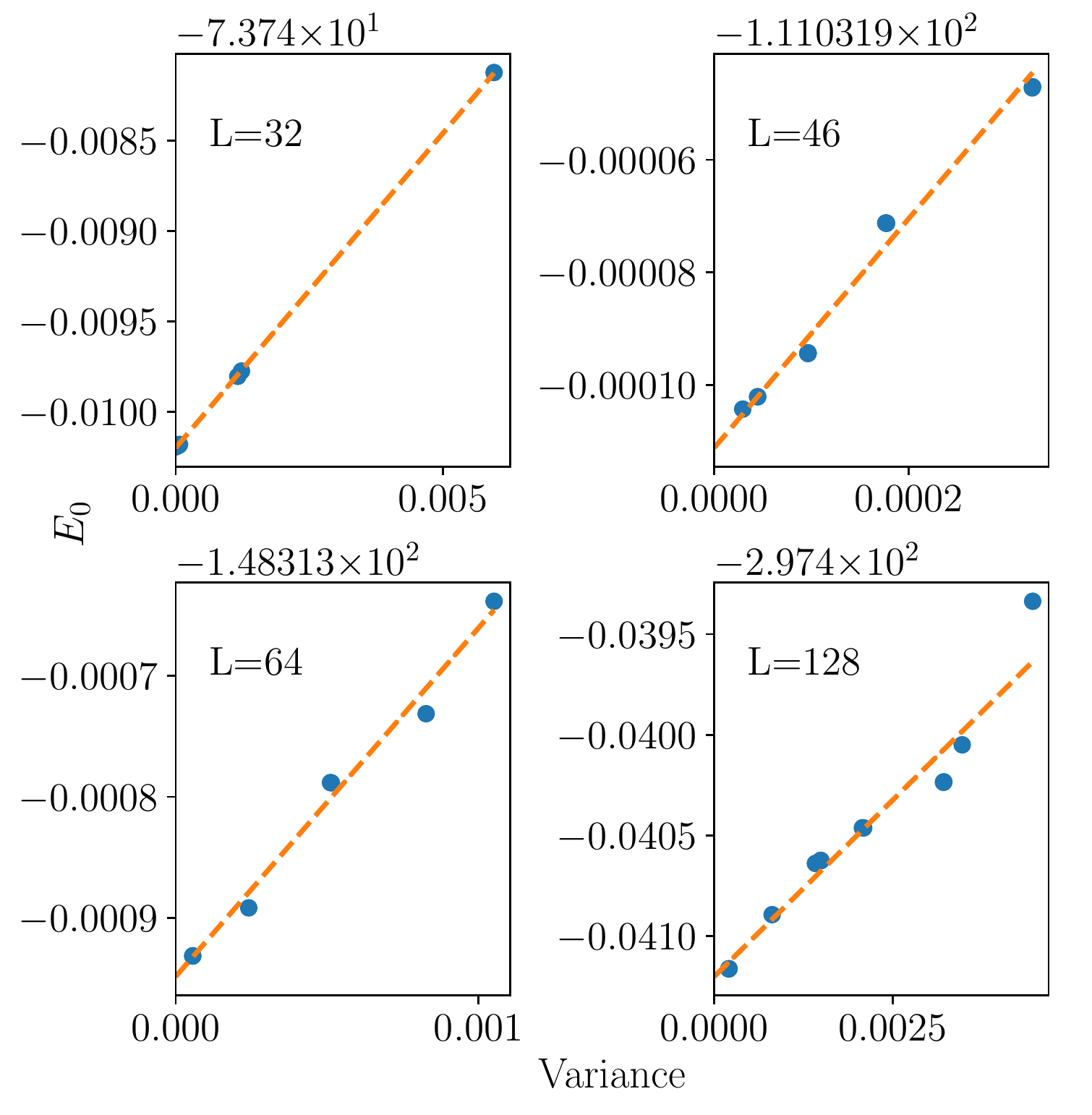}
    \caption{Variance extrapolation of the ground state ($E_0$) for various system sizes.}
    \label{fig:varience_E0}
\end{figure}
\begin{figure}
    \centering
    \includegraphics[width=0.8\columnwidth]{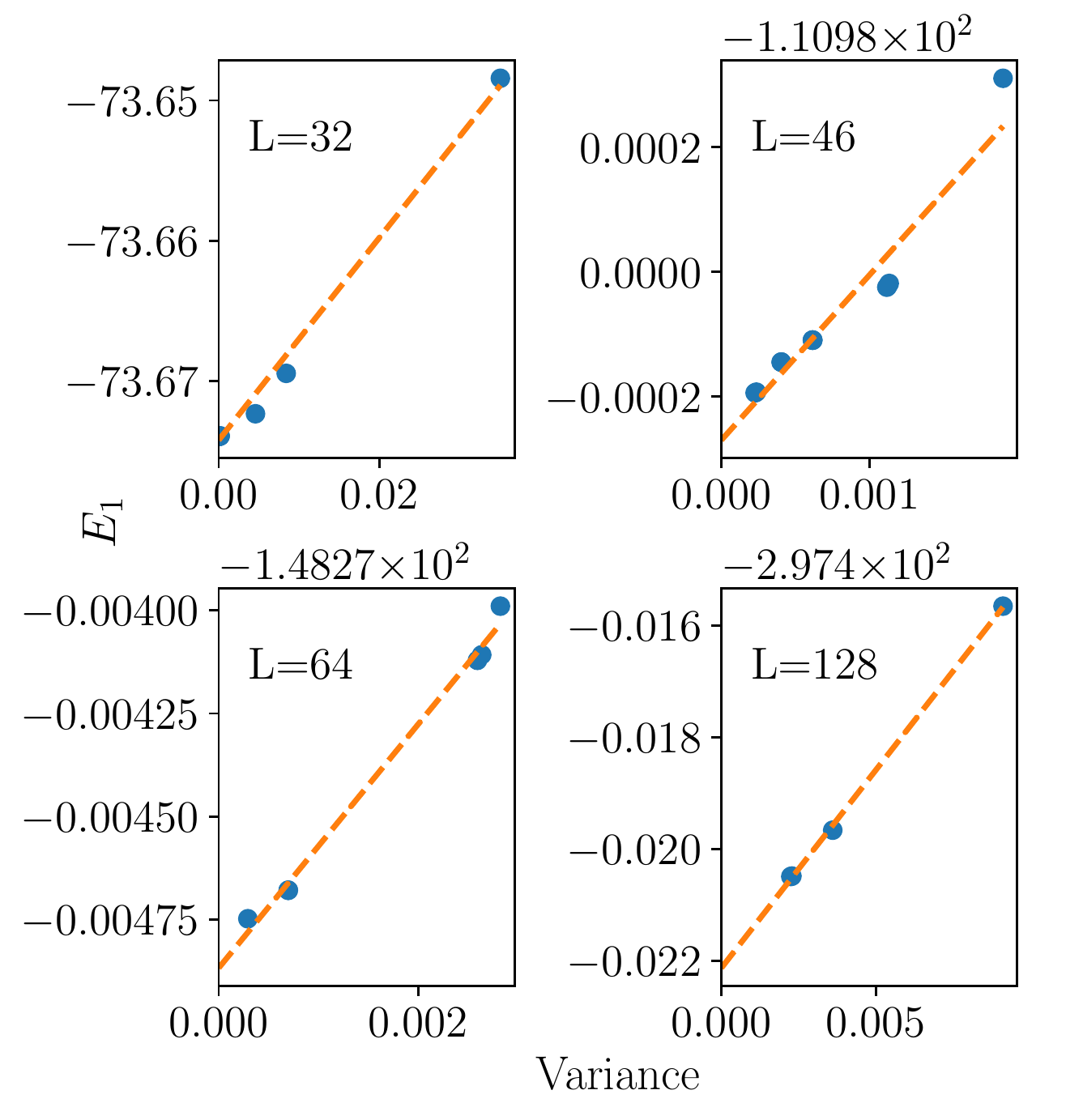}
    \caption{Variance extrapolation of the first excited state ($E_1$) for various system sizes.}
    \label{fig:varience_E1}
\end{figure}
\section{Not-Abelian Bosonization Analysis of the Kondo-Heisenberg Model}
\label{sec:NonAbel}
Here, we will now study the the problem of the Kondo-Heisenberg model using non-Abelian bosonization. There are three currents to study when considering the Kondo-Heisenberg model. A $U(1)_2$ current describing the charge degrees of freedom of the 1DEG, a $SU(2)_1$ current describing the spin degrees of freedom of the 1DEG, and a second $SU(2)_1$ current describing the spin degrees of freedom of the spin chain of the Kondo-Heisenberg, leading to a total of  current structure of $U(1)_2\times SU(2)_1\times SU(2)_1$. 

The Hamiltonian for the charge degrees of freedom is given by
\beq
\mathcal{H}_{\text{c}} &=& \frac{v_c}{2} [\frac{1}{K_c} (\partial_t \theta_{c})^2+ K_c (\partial_x \phi_{c})^2].
\label{eq:NABChargeA}
\eeq
These degrees of freedom are gapless and do not couple to the spin degrees of freedom. 

It is known that the spin currents of this model can be expressed as a $SU(2)_1\times SU(2)_1$ Wess-Zumino-Witten model.\cite{Affleck-1988} The spin currents, $\bm{J}_{e(h),R(L)}$, of the $SU(2)_1\times SU(2)_1$ WZW model are defined as 
\begin{equation}
\begin{split}
{J}^a_{e,R} = -\frac{i}{2\pi}\text{tr}(\partial_{z} g_{e} g^{-1}_{e} \tau^a), \phantom{=} {J}^a_{e,L} = \frac{i}{2\pi}\text{tr}(g^{-1}_{e}\partial_{\bar{z}} g_{e}  \tau^a),
\label{eq:curGDefA}
\end{split}
\end{equation}
and similar for $e \leftrightarrow h$. Here, $g_{e/h}$ is a $SU(2)$ matrix valued field, $\partial_{z}$ $\partial_{\bar{z}}$ are the derivative with respect to the holomophic and anti-holophomic coordinates ($t \mp i x$), and $\tau^a$ are the Pauli matrices. The OPEs for the spin currents currents are given by
\beq
\nonumber J^a_{e,R}(z)J^b_{e,R}(w) \sim  \frac{1}{(z-w)^2} \delta_{ab} +\frac{i}{(z-w)}\epsilon_{abc} J^c_{e,R}(w),\\
\label{eq:NABOPE1A}
\eeq
and similarly for $e \leftrightarrow h$ and $R \leftrightarrow L$. The lattice spins of the system are defined as 
\begin{equation}
\begin{split}
\frac{\bm{S}_{j,e}}{a} = &\frac{1}{2\pi} [\bm{J}_{e,R} (x)+\bm{J}_{e,L} (x)] + e^{i2k_f x} \Theta_e \text{tr}(g_e\bm{\tau})\\
\frac{\bm{S}_{j,h}}{a} = &\frac{1}{2\pi} [\bm{J}_{h,R} (x)+\bm{J}_{h,L} (x)] + (-1)^{x/a} \Theta_h \text{tr}(g_h\bm{\tau}),
\end{split}
\label{eq:NASpinDefA}
\end{equation}
where $\Theta_{e/h}$ are non-universal constants. The factor of $e^{i2k_f x}$ due to the doping of the electron degrees of freedom.

Using the Sugawara construction and ignoring irrelevant operators, the Hamiltonian for the spin degrees of freedom of the doped Kondo Heisenberg model is given by
\beq
\nonumber \mathcal{H}_{\text{s}} = &&\frac{2\pi v_{s,e}}{3} \bm{J}_{e,R} \bm{J}_{e,R} + \frac{2\pi v_{s,h}}{3} \bm{J}_{h,R} \bm{J}_{h,R} \\
\nonumber &-& g_{s1} [\bm{J}_{e,R} \bm{J}_{e,L} + \bm{J}_{h,R} \bm{J}_{h,L} ]\\ &-& g_{s2} [\bm{J}_{e,R} \bm{J}_{h,L} + \bm{J}_{h,R} \bm{J}_{e,L} ].
\label{eq:NABSpin1A}
\eeq
We note here that this model has a discrete symmetry that sends $(g_e,g_h) \rightarrow (-g_e,-g_h)$. If the electrons were at half filling ($k_f = \pi/2$), we would also be able to include the term $\sum_a \text{tr}(g_e\tau^a)\text{tr}(g_h\tau^a)$. However, due to the electron doping this term oscillates as $e^{i (2k_f + \pi) x}$, and is thereby irrelevant.

We will now determine the RG flow for $\mathcal{H}_{\text{s}}$. To do this, it will be useful to introduce new variables
\beq
\bm{J}_{\pm,R} =  \bm{J}_{e,R} \pm \bm{J}_{h,R} \\
\bm{J}_{\pm,L} =  \bm{J}_{e,L} \pm \bm{J}_{h,L}.
\label{eq:NABCurr1A}
\eeq
Here the $\bm{J}_{+}$ fields describe the $SU(2)_2$ currents, and $\bm{J}_{-}$ describe the remaining $SU(2)_1 \times SU(2)_1/SU(2)_2$ currents. In terms of these fields, the the spin Hamiltonian $H_\text{s}$ becomes (setting $v_{s,t} = v_{s,b} = v_s$)
\beq
\nonumber \mathcal{H}_{\text{s}} = &&\frac{2\pi v_s}{6} [\bm{J}_{+,R} \bm{J}_{+,R} +\bm{J}_{-,R} \bm{J}_{-,R}]\\
\nonumber &-& g_{+} \bm{J}_{+,R} \bm{J}_{+,L}  - g_{-} \bm{J}_{-,R} \bm{J}_{-,L}.
\label{eq:NABSpin2A}
\eeq
where $g_{\pm} = (g_{s1} \pm g_{s2})/2$. The OPE for the $\bm{J}_{\pm}$ fields are 
\beq
\nonumber J^a_{+,R}(z)J^b_{+,R}(w) &\sim &  \frac{2}{(z-w)^2} \delta_{ab} +\frac{i}{(z-w)}\epsilon_{abc} J^c_{+,R}(w)\\\nonumber 
J^a_{-,R}(z)J^b_{-,R}(w)  &\sim &  \frac{2}{(z-w)^2} \delta_{ab} +\frac{i}{(z-w)}\epsilon_{abc} J^c_{+,R}(w)\\\nonumber 
J^a_{-,R}(z)J^b_{+,R}(w) &\sim &  \frac{2}{(z-w)^2} \delta_{ab} +\frac{i}{(z-w)}\epsilon_{abc} J^c_{-,R}(w)\\\nonumber 
J^a_{+,R}(z)J^b_{-,R}(w) &\sim &  \frac{2}{(z-w)^2} \delta_{ab} +\frac{i}{(z-w)}\epsilon_{abc} J^c_{-,R}(w),\\
\label{eq:NABOPE2A}
\eeq
and similar for $L \leftrightarrow R$. Using these OPEs for the $\bm{J}_{\pm}$ fields, we have the beta functions
\beq
\nonumber \beta(g_+) &=& -\frac{2}{\pi}(g_+^2 + g_-^2)\\
\beta(g_-) &=& -\frac{4}{\pi}(g_+g_-).
\label{eq:NABBeta1A}
\eeq
Let us examine the $\beta$ functions near $(g_+,g_-) = (0,0)$. For $g_- \neq 0$ or $g_+ < 0$, $g_+$ flows to $-\infty$. For $g_+ < 0$, we can rewrite the $\beta(g_-)$ as 
\beq
\beta(g_-  ) = \frac{4}{\pi}|g_+| g_-.
\label{eq:NABBeta2A}
\eeq
Rewriting $g_-$  as $\pm |g_-|$, we have that 
\beq
\beta(|g_- |) = \frac{4}{\pi}|g_+| |g_-|.
\label{eq:NABBeta3A}
\eeq
So for $g_+ < 0$, $g_- > 0 $, $g_-$ flows to $\infty$ and for $g_+ < 0$, $g_- < 0 $, $g_-$ flows to $- \infty$. 
Using this, we can identify the fixed points $(g_+,g_-) = (0,0)$, $(-\infty,\infty)$, $(-\infty,-\infty)$, and $(-\infty,0)$.

\section{Fermion Number in the $SU(2)_1 \times SU(2)_1$ WZW model}
\label{sec:FermNum}
Let us consider the $SU(2)_1 \times SU(2)_1$ WZW model defined on a ring. This is equivalent to defining the WZW model on the complex plane where the radial direction is time, and the polar angle is space. We can express the $SU(2)$ currents in terms of Majoranas using 
\beq
J^a_{e,R/L} = \frac{i}{2}\left(\frac{\epsilon^{abc}}{2}\eta^b_{R/L}\eta^c_{R/L} + \eta^0_{R/L}\eta^a_{R/L}\right),\\
J^a_{h,R/L} = \frac{i}{2}\left(\frac{\epsilon^{abc}}{2}\eta^b_{R/L}\eta^c_{R/L} - \eta^0_{R/L}\eta^a_{R/L}\right),
\label{eq:MFDef3}
\eeq

Let us now define the following charge operator:
\beq
\nonumber N_f = \frac{2}{2\pi i} \oint &dz &(J^3_{e,R}(z) + J^3_{e,L}(\bar{z}))\\\nonumber = \frac{1}{2\pi i} \oint &dz &(\eta_{1,R}(z)\eta_{2,R}(z) + \eta_{0,R}(z)\eta_{3,R}(z)\\ &+& \eta_{1,L}(\bar{z})\eta_{2,L}(\bar{z}) + \eta_{0,L}(\bar{z})\eta_{3,L}(\bar{z})),
\eeq
where the contour integral is over a circle of constant radius in the complex plane, i.e., a constant time slice. The charge $q_A$ of a field $A(w,\bar{w})$ is given by
\beq
[N_f,A(w,\bar{w})] = q_A A(w,\bar{w}),
\eeq
were $[...]$ is the radially ordered commutator. We find that the $\bm{J}_{e,R}$ currents have the following charges:
\beq
[N_f, J^3_{e,R}(w)] &=&   0
\label{eq:ComFirst}
\eeq
\beq
[N_f, J^{\pm}_{e,R}(w)] &=&  \pm 2 J^{\pm}_{e,R}(w).
\eeq
The charges of  the $\bm{J}_{e,L}$ are identical. The charge of components of the matrix valued $WZW$ field $g_e$ are:
\beq
[N_f, g_e(w,\bar{w})_{00}] &=& 2 g_e(w,\bar{w})_{00}
\eeq
\beq
[N_f, g_e(w,\bar{w})_{01}] &=& [N_f, g_e(w,\bar{w})_{10}] = 0
\eeq
\beq
[N_f, g_e(w,\bar{w})_{11}] &=& -2 g_e(w,\bar{w})_{11},
\eeq
where $g_e(w,\bar{w})_{ij}$ are the components of the matrix valued $WZW$ field $g_e$. The charges of the (sum of) Majoranas are:
\begin{equation}
\begin{split}
[N_f, \eta_{1,R}(w)\pm i\eta_{2,R}(w)]  = \pm (\eta_{1,R}(w)\pm i\eta_{2,R}(w))
\end{split}
\end{equation}
\begin{equation}
\begin{split}
[N_f, \eta_{0,R}(w)\pm i\eta_{3,R}(w)] = \pm (\eta_{0,R}(w)\pm i\eta_{3,R}(w)).
\end{split}
\end{equation}
The charges of the left handed Majoranas are the same. Additionally,
\begin{equation}
\begin{split}
[N_f, J^a_{h,R}(w)] = [N_f, J^a_{h,L}(\bar{w})] = [N_f, g_{h}(w,\bar{w})_{ij}] = 0,
\label{eq:ComTriv}
\end{split}
\end{equation}
since all the OPEs disappear. From this we can conclude that fields $\bm{J}_{e/h,R/L}$ and $g_{e,h}$ all have charge $0$ mod$(2)$. The Majorana fields $\eta_{\mu, R/L}$ have charge $1$ mod$(2)$. As such $\bm{J}_{e/h,R/L}$ and $g_{e,h}$ all have even charge parity, $(-1)^{N_f}$, while $\eta_{\mu R/L}$ have odd charge parity.

We can also find the charges of the individual modes of spin currents $\bm{J}_{e/h,R/L}$ and Majorana currents $\eta_{\mu, R/L}$ analogously. The modes in radial quantization are respectively
\beq
J^a_{n,e,R} = \oint \frac{dw}{2\pi i} w^n J_{e,R}(w)
\label{eq:ModeC}
\eeq 
\beq
\eta_{n,\mu,R} = \oint \frac{dw}{2\pi} w^{n-1/2} \eta_{\mu,R}(w),
\label{eq:ModeM}
\eeq 
and similar for $e \rightarrow h$ and $R \rightarrow L$. Combining Eq. \ref{eq:ModeC} and \ref{eq:ModeM} with Eq. \ref{eq:ComFirst}-\ref{eq:ComTriv}, we find that all modes $J^a_{n}$ have even charge $0$ mod$(2)$ and all modes $\eta_{n}$ have charge $1$ mod$(2)$ as expected.

Let us now consider the charge of various states in the Hilbert space of the $SU(2)_1 \times SU(2)_1$ WZW model. It is known that the Hilbert space of a CFT can be divided into Verma modules. The Verma modules are built off of a highest weight state. In the $SU(2)_1 \times SU(2)_1$ WZW model there are four highest weight states. First there is the trivial vacuum state which we will label $\ket{0}$. Second there are the highest weight states that correspond to inserting a primary field. For $SU(2)_1 \times SU(2)_1$ WZW model in radial quantization, the primary fields that are inserted are $g_{e}(0,0)_{ij}$, $g_{h}(0,0)_{ij}$ and their product $g_{e}(0,0)_{ij}g_{h}(0,0)_{kl}$. We will label the corresponding highest weight states as $g_{e,ij}\ket{0}$ and $g_{h,ij}\ket{0}$ and $g_{e,ij}g_{h,kl}\ket{0}$. The descendant states of these highest weight states are created by acting on the highest weight states with the operators $J^a_{-n,e/h,R/L}$. 

Let us now consider the parity of a state in the Hilbert space. A general state built off the vacuum highest weight state $\ket{0}$ can be written as 
\beq
J^a_{-n_1,e/h, L/R}J^b_{-n_2,e/h, L/R}... \ket{0}.
\label{eq:State1}
\eeq
From our earlier analysis we know that the modes $J^a_{-n_1,e/h, L/R}$ have charge $0$ mod$(2)$. Since the vacuum has charge 0 by definition, we can conclude that all states built off the vacuum have even charge i.e. $(-1)^{N_f} = 1$ for all states in Eq. \ref{eq:State1}. 

We will now consider the other states in the Hilbert space that are built off the $g_{e,ij}\ket{0}$ and $g_{h,ij}\ket{0}$ and $g_{e,ij}g_{h,kl}\ket{0}$ highest weight states. In general, these states can be written as 
\beq
\nonumber J^a_{-n_1,e/h, L/R}J^b_{-n_2,e/h, L/R}... g_{e,ij}\ket{0}\phantom{.}\\\nonumber
J^a_{-n_1,e/h, L/R}J^b_{-n_2,e/h, L/R}... g_{h,ij}\ket{0}\phantom{.}\\
J^a_{-n_1,e/h, L/R}J^b_{-n_2,e/h, L/R}... g_{e,ij}g_{h,kl}\ket{0}.
\label{eq:State2}
\eeq
As before, we know that the modes $J^a_{-n_1,e/h, L/R}$ have charge $0$ mod$(2)$. Our earlier analysis has also shown that $g_{e}(0,0)_{ij}$, $g_{h}(0,0)_{ij}$ and their product $g_{e}(0,0)_{ij}g_{h}(0,0)_{kl}$ all have charge $0$ mod$(2)$. Because of this, $(-1)^{N_f} = 1$ for all state in Eq. \ref{eq:State2}. From this we can conclude that $(-1)^{N_f} = 1$ for all states in the Hilbert space of the $SU(2)\times SU(2)$ WZW model. 

Let us now consider acting on a given even charge parity state $\ket{\psi}$ with a single Majorana fermion mode $\eta_{-n,\mu,R/L}$. 
\beq
\eta_{-n,\mu,R/L}\ket{\psi}.
\label{eq:State3}
\eeq
From our earlier result, we know that $\eta_{-n,\mu,R/L}$ has charge $1$ mod$(2)$. Since the state $\ket{\psi}$ has $(-1)^{N_f} = 1$, the state in Eq. \ref{eq:State3} has $(-1)^{N_f} = -1$. However, we know that all states in Hilbert space of the $SU(2)\times SU(2)$ WZW model have $(-1)^{N_f} = 1$. So the state in Eq. \ref{eq:State3} cannot be a physical state of the $SU(2)\times SU(2)$ WZW model. We can also consider acting the state $\ket{\psi}$ with two Majorana fermion modes $\eta_{-n,\mu,R/L}$ and $\eta_{-n',\nu,R/L}$,
\beq
\eta_{-n,\mu,R/L}\eta_{-n',\nu,R/L}\ket{\psi}.
\label{eq:State4}
\eeq
Since each of the modes have charge $1$ mod$(2)$, the state in Eq. \ref{eq:State4} has $(-1)^{N_f} = 1$. So this can be a physical state in the $SU(2)\times SU(2)$ WZW model. We can thereby conclude that a single Majorana mode operator is not a physical operator in the $SU(2)\times SU(2)$ WZW model. In other words there are no single fermions modes in the spectrum of the $SU(2)\times SU(2)$ WZW model. The fermions only occur as bilinears. 

\section{Continuum limit and Abelian Bosonization}
\label{sec:ContLim}
Here, we will now discuss the continuum limit of the lattice model using Abelian bosonization. In the low energy limit, the fermions and spins can be expressed in terms of continuum current operators:
\beq
\nonumber \frac{1}{\sqrt{a}}c_{j,\sigma} &\rightarrow & R_{\sigma}(x)e^{ik_f x} + L_{\sigma,t}(x)e^{-ik_f x}\\
\frac{\bm{S}_{j,h}}{a}  &\rightarrow & \bm{J}_{h,R}(x) + \bm{J}_{h,L}(x) + (-1)^{x/a} \bm{N}_h(x).
\label{eq:Bosons1}
\eeq
Here, $R_{\sigma}$ and $L_{\sigma}$ are the right and left moving components of the electron fields, $\bm{J}_{h,R}$ and $\bm{J}_{h,L}$are the slowly varying components of the spin field, $\bm{N}_h$ is the rapidly oscillating (Néel) component of the spin field, and $x = ja$ where $a$ is the lattice spacing.

The right and left moving continuum fields can be bosonized using the following identifications
\beq
\nonumber R_{\sigma} &=& :\frac{1}{\sqrt{2\pi a}} e^{-i \sqrt{2\pi} [\phi_{\sigma} + \sigma \theta_{\sigma}]}:\\
\nonumber L_{\sigma} &=& :\frac{1}{\sqrt{2\pi a}} e^{i  \sqrt{2\pi}[\phi_{\sigma} + \sigma \theta_{\sigma}]}:\\
\nonumber J^z_{h,R} &=&\frac{1}{\sqrt{2\pi}} \partial_x [\tilde{\phi}_s - \tilde{\theta}_s]\\\nonumber 
J^z_{h,L} &=& \frac{1}{\sqrt{2\pi}} \partial_x [\tilde{\phi}_s + \tilde{\theta}_s]\\\nonumber
J^\pm_{h,R} &=& :\frac{1}{2\pi a} e^{\mp i  \sqrt{2\pi}[\tilde{\phi}_s - \tilde{\theta}_s]}:\\
J^\pm_{h,L} &=& :\frac{1}{2\pi a} e^{\pm i  \sqrt{2\pi}[\tilde{\phi}_s + \tilde{\theta}_s]}:
\label{eq:Bosons2}
\eeq
where $\phi_{\sigma}$ and $\theta_{\sigma}$ are the field and dual field of the electrons, $\tilde{\phi}_s$ and $\tilde{\theta}_s$ are the field and dual field of the spins, and $:...:$ indicates normal ordering of the exponential. From here on we will leave the normal ordering implicit. In this definition, we note that the the spin fields are defined such that $\tilde{\phi_s} \equiv \tilde{\phi_s} +\sqrt{2\pi}$.

It will be useful to decompose the field and dual fields into spin and charge degrees of freedom using
\beq
\nonumber \phi_{c}&=& \frac{1}{\sqrt{2}} (\phi_{\uparrow} +\phi_{\downarrow})\\
\phi_{s} &=& \frac{1}{\sqrt{2}} (\phi_{\uparrow} -\phi_{\downarrow}),
\label{eq:Boson3}
\eeq
and similarly for the $\theta$ fields. 

The dominant interactions will be between the the 1DEG spins and the Heisenberg spins. In terms of the spin currents of the 1DEG $\bm{J}_{e,R}  = \frac{1}{2}R_{\sigma} \bm{\tau}_{\sigma,\sigma'}R_{\sigma'}$ and $\bm{J}_{e,L}  = \frac{1}{2}L_{\sigma} \bm{\tau}_{\sigma,\sigma'}L_{\sigma'}$, the most general interactions consistent with the $SU(2)$ symmetries of the model are:
\beq
\nonumber H_{int} = &-& g_{s1,e} \bm{J}_{e,R} \bm{J}_{e,L} - g_{s1,s}\bm{J}_{h,R} \bm{J}_{h,L} ]\\&-& g_{s2} [\bm{J}_{e,R} \bm{J}_{h,L} + \bm{J}_{h,R} \bm{J}_{e,L} ].
\label{eq:IntCurr}
\eeq
At weak coupling, the relationship between the values of these coupling constants and those of the microscopic model Eq. \ref{eq:LatHam} are $g_{s1,e} = U$, $g_{s1,s} = J_H-6J'_H$, and $g_{s2} = -J_K$. As noted before, the microscopic model Eq. \ref{eq:LatHam} can also arise as the effective description of a two-leg Hubbard ladder \cite{jaefari2012}. The relationship between the coupling constants in Eq. \ref{eq:IntCurr} and the those of the two leg Hubbard ladder are more complex and can be found in\cite{wu2003}. 

If we set $g_{s1,e}=g_{s1,s}\equiv g_{s1}$ and define $\phi_{s,\pm} \equiv \frac{1}{\sqrt{2}}(\phi_{s} \pm \tilde{\phi}_s)$ and similarly for $\theta_{s,\pm}$, we arrive at the continuum Hamiltonian:
\begin{equation}
\begin{split}
\mathcal{H} =& \mathcal{H}_c+\mathcal{H}_s\\
\mathcal{H}_c =& \frac{v_c}{2} [ K_c(\partial_t \theta_{c,e})^2 + \frac{1}{K_c} (\partial_x \phi_{c,e})^2]\\
\mathcal{H}_s =& \sum_{\epsilon = \pm}\frac{v_{s,\epsilon}}{2} [ K_{s,\epsilon}(\partial_t \theta_{s,\epsilon})^2 + \frac{1}{K_{s,\epsilon}} (\partial_x \phi_{s,\epsilon})^2]
\\ &  +\frac{g_{s1}}{2(\pi a)^2} \cos(\sqrt{4\pi}\phi_{s,+})\cos(\sqrt{4\pi}\phi_{s,-}) \\
\phantom{=}& +  \frac{g_{s2}}{2(\pi a)^2} \cos(\sqrt{4\pi}\phi_{s,+})\cos(\sqrt{4\pi}\theta_{s,-}).
\label{eq:BosonHam1A}
\end{split}
\end{equation}
It is important to note that since $\tilde{\phi_{s}} \equiv \tilde{\phi_{s}} + \sqrt{2\pi}$, ($(\phi_{s,+},\phi_{s,-}) \equiv (\phi_{s,+} + \sqrt{\pi},\phi_{s,-}-\sqrt{\pi})$.

\section{Majorana Zero modes using Abelian Bosonization}
\label{sec:MajorAbel}
The original argument for the existence of MZMs\cite{cho2014} comes from considering a section of PDW wire ($(g_K,g_{SC}) = ( -\infty,0)$) of length $L$ that is sandwiched in between between two sections of trivial SC wire ($(g_K,g_{SC}) = ( 0,-\infty)$). So the wire is in a trivial SC state for $x<0$ and $L < x $, and a PDW phase for $0<x < L$. In the analysis of the topological features of the PDW wire, we will only be interested in the gapped spin sector of the wire (Eq. \ref{eq:BosonHam1A}), and not in the gapless charge sectors. We will also take $L$ to be much greater than the correlation length of the spin sector. 
To show the proposed MZM which are localized at $x=0$ and $x=L$, we will refermionize Eq. \ref{eq:BosonHam1A} around the $K_{s\pm} = 1$ point. Assuming that $\phi_{s+}$ is pinned to the same minimum throughout the entire system, the refermionized Hamiltonian is given by
\beq
\nonumber \mathcal{H}_s &=& -i v_s (\mathcal{R}^\dagger \partial_x \mathcal{R} - \mathcal{L}^\dagger \partial_x \mathcal{L}) \\\nonumber &\phantom{=}& + M_{USC} \mathcal{R}^\dagger \mathcal{L} + \Delta_{PDW} \mathcal{R}^\dagger \mathcal{L}^\dagger + h.c.\\\nonumber
 \mathcal{R} &\sim & e^{-i \sqrt{\pi}(\phi_{s,-}-\theta_{s,-})}\\
 \mathcal{L} &\sim & e^{i \sqrt{\pi}(\phi_{s,-}+\theta_{s,-})}
\label{eq:Referm}
\eeq
where $M_{USC} \sim g_{s1}\langle \cos(\sqrt{4\pi}\phi_{s+})\rangle $ and $\Delta_{PDW} \sim g_{s2}\langle \cos(\sqrt{4\pi}\phi_{s+})\rangle$. Fermion number is not conserved in Eq \ref{eq:Referm}, but the fermion parity given by 
\beq
(-1)^{N_f} = (-1)^{\int dx \mathcal{R}^\dagger \mathcal{R}+\mathcal{L}^\dagger\mathcal{L}} 
\label{eq:FermParity1A}
\eeq
is conserved.

 Decomposing the fermions into Majorana fermions using $\mathcal{R} = \frac{1}{\sqrt{2}}(\eta_{1,R} + i\eta_{2,R}) $, $\mathcal{L} =  \frac{1}{\sqrt{2}}(\eta_{2,L} + i\eta_{1,L})$ the potential term in Eq. \ref{eq:Referm} becomes 
 \beq
 \nonumber \mathcal{V}_s &=& (M_{USC}-\Delta_{PDW})i\eta_{1,R}\eta_{1,L}\\ &\phantom{=}& + (M_{USC}+\Delta_{PDW})i\eta_{2,R}\eta_{2,L}.
\label{eq:FermPotential} 
 \eeq
Since $M_{USC} \sim g_{s1}$ and $\Delta_{PDW} \sim g_{s2}$, $M_{USC}-\Delta_{PDW}$ changes sign when moving from the SC region to the PDW region at $x=0$ and $x= L$.  At these points there will be zero energy mode for the Majorana fermions $\eta_1 = (\eta_{1,R},\eta_{1,L})$ due to the Jackiw-Rebbi mechanism. These Majorana zero modes imply that the spin sector of the doped Kondo-Heisenberg model can be considered to be topological superconductor in class \textbf{D}, i.e. a Kiteav chain. Acting on a given state with a MZM operator changes the fermion parity of the state (Eq. \ref{eq:FermParity1A}) from $\pm 1$ to $\mp 1$. Naively this would lead to two ground states, one with fermion parity even, and one with fermion parity odd.

It is at this point that we wish to ask if the Majorana zero modes from the refermionized model Eq. \ref{eq:Referm} correspond to physical operators in the original spin model. To answer this in the Abelian bosonized framework, we first note that the bosonic fields describing the electron and Heisenberg spins are compact. This compactification means that the field $\phi_{\pm}$ are defined such that $(\phi_{s+},\phi_{s-}) \equiv (\phi_{s+} + \sqrt{\pi},\phi_{s-}-\sqrt{\pi})$  (see Appendix \ref{sec:ContLim}). Because of this, all physical operators in the theory must be invariant under sending $\phi_{s\pm} \rightarrow \phi_{s\pm} \pm \sqrt{\pi}$ simultaneously. However, if in the refermionization in Eq.\ref{eq:Referm}, we note that the fermions $\mathcal{R}$ and $\mathcal{L}$ are not not invariant under this transformation, but instead transform as $\mathcal{R} \rightarrow -\mathcal{R}$ and $\mathcal{L} \rightarrow -\mathcal{L}$. 

Let us now consider the situation where there is a boundary between a section of PDW wire and a section of trivial SC wire. As noted before, one would expect there to be a pair of Majorana zero modes at either ends of the PDW wire. Let us consider the ground state of the system $\ket{0}$, that must be invariant under the transformation $\phi_{s\pm} \rightarrow \phi_{s\pm} \pm \sqrt{\pi}$. Furthmore, the ground state will have a well defined fermion parity (Eq. \ref{eq:FermParity1A}), that we will take to be equal $+1$. If one of the zero mode $\eta_1$ is physical, the state $\eta_1\ket{0}$, will be a degenerate ground state with fermion parity $-1$. However, as noted before, the zero mode $\eta_1$ is not a physical state, since under $\phi_{s\pm} \rightarrow \phi_{s\pm} \pm \sqrt{\pi}$, $\eta_1 \rightarrow -\eta_1$. So $\eta_1\ket{0}$ cannot be a physical state. From this, we can also conclude that only products of an even number of Majorana fermion operators lead to physical states. This means that all physical states will have fermion parity  $+1$. From our earlier logic we can then confirm that there is not ground state degeneracy, and by extension no Majorana zero modes.

\section{$\mathbb{Z}_2$ order parameter}
\label{sec:OrderParam}
Here we discuss the a $\mathbb{Z}_2$ order parameter and the associated symmetry breaking that occurs between the trivial SC and PDW phases of the doped Kondo Heisenberg model. We expect this symmetry breaking to occur because the phase transition between the two phases in the Ising universality class. The $\mathbb{Z}_2$ symmetry that we will need to consider sends $(g_e,g_h) \rightarrow (-g_e,-g_h)$.

To approach the problem of symmetry breaking, it will be useful to consider this problem with Abelian bosonization instead of non-Abelian bosonization, since in the former case, the order parameters can be read off by using a semi-classical analysis. In Abelian bosonization, the WZW fields $g_e$ and $g_h$ can be written as 
\begin{equation}
\begin{split}
g_e = \begin{bmatrix}
e^{i\sqrt{2\pi} \phi_s } & e^{-i\sqrt{2\pi} \theta_s }\\
-e^{i\sqrt{2\pi} \theta_s } & e^{-i\sqrt{2\pi} \phi_s }\end{bmatrix}, \phantom{=}g_h = \begin{bmatrix}e^{i\sqrt{2\pi} \tilde{\phi}_s } & e^{-i\sqrt{2\pi} \tilde{\theta}_s }\\
-e^{i\sqrt{2\pi} \tilde{\theta}_s } & e^{-i\sqrt{2\pi} \tilde{\phi}_s }
\end{bmatrix}.
\label{eq:gDef}
\end{split}
\end{equation}
We note that combining Eq. \ref{eq:gDef} and \ref{eq:curGDefA} reproduces Eq. \ref{eq:Bosons2}. The order parameters we are interested in will be $\text{tr}(g_e) = 2\cos (\sqrt{2\pi} \phi_{s})$, and $\text{tr}(g_h) = 2\cos (\sqrt{2\pi} \tilde{\phi}_{s})$, as well as their product $\text{tr}(g_e)\text{tr}(g_h)$. Clearly $\text{tr}(g_e)$ and $\text{tr}(g_h)$ are odd under $(g_e,g_h) \rightarrow (-g_e,-g_h)$ but their product is not. 

Let us now determine when these order parameters have expectation values. In the trivial SC phase, $\langle \phi_{s+} \rangle =\langle \phi_{s-} \rangle = 0,\sqrt{\pi/2}$. Using that  $\phi_{s\pm} = \frac{1}{\sqrt{2}}(\phi_{s} \pm \tilde{\phi}_{s} )$, we can determine that both $\langle \text{tr}(g_e)\rangle = \pm 2$ and $\langle \text{tr}(g_e)\rangle = \pm 2$, and so the $\mathbb{Z}_2$ symmetry is broken. In the PDW phase $\langle \phi_{s+} \rangle =\langle \theta_{s-} \rangle = 0,\sqrt{\pi/2}$. In this phase, neither $\text{tr}(g_e)$ or $\text{tr}(g_e)$ have expectation values, but their product does have an expectation value $\langle \text{tr}(g_e)\text{tr}(g_h)\rangle = \pm 2$. So, the $(g_e,g_h) \rightarrow (-g_e,-g_h)$ symmetry is unbroken. We can thereby identify the phase transition between the PDW and trivial SC phase with breaking the $(g_e,g_h) \rightarrow (-g_e,-g_h)$ symmetry.

\bibliography{PDW_Not_Topo.bib}

\begin{thebibliography}{38}%
\makeatletter
\providecommand \@ifxundefined [1]{%
 \@ifx{#1\undefined}
}%
\providecommand \@ifnum [1]{%
 \ifnum #1\expandafter \@firstoftwo
 \else \expandafter \@secondoftwo
 \fi
}%
\providecommand \@ifx [1]{%
 \ifx #1\expandafter \@firstoftwo
 \else \expandafter \@secondoftwo
 \fi
}%
\providecommand \natexlab [1]{#1}%
\providecommand \enquote  [1]{``#1''}%
\providecommand \bibnamefont  [1]{#1}%
\providecommand \bibfnamefont [1]{#1}%
\providecommand \citenamefont [1]{#1}%
\providecommand \href@noop [0]{\@secondoftwo}%
\providecommand \href [0]{\begingroup \@sanitize@url \@href}%
\providecommand \@href[1]{\@@startlink{#1}\@@href}%
\providecommand \@@href[1]{\endgroup#1\@@endlink}%
\providecommand \@sanitize@url [0]{\catcode `\\12\catcode `\$12\catcode
  `\&12\catcode `\#12\catcode `\^12\catcode `\_12\catcode `\%12\relax}%
\providecommand \@@startlink[1]{}%
\providecommand \@@endlink[0]{}%
\providecommand \url  [0]{\begingroup\@sanitize@url \@url }%
\providecommand \@url [1]{\endgroup\@href {#1}{\urlprefix }}%
\providecommand \urlprefix  [0]{URL }%
\providecommand \Eprint [0]{\href }%
\providecommand \doibase [0]{http://dx.doi.org/}%
\providecommand \selectlanguage [0]{\@gobble}%
\providecommand \bibinfo  [0]{\@secondoftwo}%
\providecommand \bibfield  [0]{\@secondoftwo}%
\providecommand \translation [1]{[#1]}%
\providecommand \BibitemOpen [0]{}%
\providecommand \bibitemStop [0]{}%
\providecommand \bibitemNoStop [0]{.\EOS\space}%
\providecommand \EOS [0]{\spacefactor3000\relax}%
\providecommand \BibitemShut  [1]{\csname bibitem#1\endcsname}%
\let\auto@bib@innerbib\@empty
\bibitem [{\citenamefont {Li}\ \emph {et~al.}(2007)\citenamefont {Li},
  \citenamefont {H{\"u}cker}, \citenamefont {Gu}, \citenamefont {Tsvelik},\
  and\ \citenamefont {Tranquada}}]{Li-2007}%
  \BibitemOpen
  \bibfield  {author} {\bibinfo {author} {\bibfnamefont {Q.}~\bibnamefont
  {Li}}, \bibinfo {author} {\bibfnamefont {M.}~\bibnamefont {H{\"u}cker}},
  \bibinfo {author} {\bibfnamefont {G.~D.}\ \bibnamefont {Gu}}, \bibinfo
  {author} {\bibfnamefont {A.~M.}\ \bibnamefont {Tsvelik}}, \ and\ \bibinfo
  {author} {\bibfnamefont {J.~M.}\ \bibnamefont {Tranquada}},\ }\href {\doibase
  10.1103/PhysRevLett.99.067001} {\bibfield  {journal} {\bibinfo  {journal}
  {Phys. Rev. Lett.}\ }\textbf {\bibinfo {volume} {99}},\ \bibinfo {pages}
  {067001} (\bibinfo {year} {2007})}\BibitemShut {NoStop}%
\bibitem [{\citenamefont {Berg}\ \emph {et~al.}(2007)\citenamefont {Berg},
  \citenamefont {Fradkin}, \citenamefont {Kim}, \citenamefont {Kivelson},
  \citenamefont {Oganesyan}, \citenamefont {Tranquada},\ and\ \citenamefont
  {Zhang}}]{Berg-2007}%
  \BibitemOpen
  \bibfield  {author} {\bibinfo {author} {\bibfnamefont {E.}~\bibnamefont
  {Berg}}, \bibinfo {author} {\bibfnamefont {E.}~\bibnamefont {Fradkin}},
  \bibinfo {author} {\bibfnamefont {E.-A.}\ \bibnamefont {Kim}}, \bibinfo
  {author} {\bibfnamefont {S.~A.}\ \bibnamefont {Kivelson}}, \bibinfo {author}
  {\bibfnamefont {V.}~\bibnamefont {Oganesyan}}, \bibinfo {author}
  {\bibfnamefont {J.~M.}\ \bibnamefont {Tranquada}}, \ and\ \bibinfo {author}
  {\bibfnamefont {S.~C.}\ \bibnamefont {Zhang}},\ }\href {\doibase
  10.1103/PhysRevLett.99.127003} {\bibfield  {journal} {\bibinfo  {journal}
  {Phys. Rev. Lett.}\ }\textbf {\bibinfo {volume} {99}},\ \bibinfo {pages}
  {127003} (\bibinfo {year} {2007})}\BibitemShut {NoStop}%
\bibitem [{\citenamefont {Berg}\ \emph {et~al.}(2009)\citenamefont {Berg},
  \citenamefont {Fradkin}, \citenamefont {Kivelson},\ and\ \citenamefont
  {Tranquada}}]{Berg-2009}%
  \BibitemOpen
  \bibfield  {author} {\bibinfo {author} {\bibfnamefont {E.}~\bibnamefont
  {Berg}}, \bibinfo {author} {\bibfnamefont {E.}~\bibnamefont {Fradkin}},
  \bibinfo {author} {\bibfnamefont {S.~A.}\ \bibnamefont {Kivelson}}, \ and\
  \bibinfo {author} {\bibfnamefont {J.~M.}\ \bibnamefont {Tranquada}},\ }\href
  {\doibase 10.1088/1367-2630/11/11/115004} {\bibfield  {journal} {\bibinfo
  {journal} {New J. Phys.}\ }\textbf {\bibinfo {volume} {11}},\ \bibinfo
  {pages} {115004} (\bibinfo {year} {2009})}\BibitemShut {NoStop}%
\bibitem [{\citenamefont {Edkins}\ \emph {et~al.}(2019)\citenamefont {Edkins},
  \citenamefont {Kostin}, \citenamefont {Fujita}, \citenamefont {Mackenzie},
  \citenamefont {Eisaki}, \citenamefont {Uchida}, \citenamefont {Lawler},
  \citenamefont {Kim}, \citenamefont {Davis},\ and\ \citenamefont
  {Hamidian}}]{Edkins-2018}%
  \BibitemOpen
  \bibfield  {author} {\bibinfo {author} {\bibfnamefont {S.~D.}\ \bibnamefont
  {Edkins}}, \bibinfo {author} {\bibfnamefont {A.}~\bibnamefont {Kostin}},
  \bibinfo {author} {\bibfnamefont {K.}~\bibnamefont {Fujita}}, \bibinfo
  {author} {\bibfnamefont {A.~P.}\ \bibnamefont {Mackenzie}}, \bibinfo {author}
  {\bibfnamefont {H.}~\bibnamefont {Eisaki}}, \bibinfo {author} {\bibfnamefont
  {S.}~\bibnamefont {Uchida}}, \bibinfo {author} {\bibfnamefont {M.~J.}\
  \bibnamefont {Lawler}}, \bibinfo {author} {\bibfnamefont {E.-A.}\
  \bibnamefont {Kim}}, \bibinfo {author} {\bibfnamefont {J.}~\bibnamefont
  {Davis}}, \ and\ \bibinfo {author} {\bibfnamefont {M.~H.}\ \bibnamefont
  {Hamidian}},\ }\href {\doibase 10.1126/science.aat1773} {\bibfield  {journal}
  {\bibinfo  {journal} {Science}\ }\textbf {\bibinfo {volume} {364}},\ \bibinfo
  {pages} {976} (\bibinfo {year} {2019})}\BibitemShut {NoStop}%
\bibitem [{\citenamefont {Fulde}\ and\ \citenamefont
  {Ferrell}(1964)}]{Fulde-1964}%
  \BibitemOpen
  \bibfield  {author} {\bibinfo {author} {\bibfnamefont {P.}~\bibnamefont
  {Fulde}}\ and\ \bibinfo {author} {\bibfnamefont {R.~A.}\ \bibnamefont
  {Ferrell}},\ }\href {\doibase 10.1103/PhysRev.135.A550} {\bibfield  {journal}
  {\bibinfo  {journal} {Phys. Rev.}\ }\textbf {\bibinfo {volume} {135}},\
  \bibinfo {pages} {A550} (\bibinfo {year} {1964})}\BibitemShut {NoStop}%
\bibitem [{\citenamefont {Larkin}\ and\ \citenamefont
  {Ovchinnikov}(1964)}]{Larkin-1964}%
  \BibitemOpen
  \bibfield  {author} {\bibinfo {author} {\bibfnamefont {A.~I.}\ \bibnamefont
  {Larkin}}\ and\ \bibinfo {author} {\bibfnamefont {Y.~N.}\ \bibnamefont
  {Ovchinnikov}},\ }\href@noop {} {\bibfield  {journal} {\bibinfo  {journal}
  {Zh. Eksp. Teor. Fiz.}\ }\textbf {\bibinfo {volume} {47}},\ \bibinfo {pages}
  {1136} (\bibinfo {year} {1964})},\ \bibinfo {note} {[Sov. Phys. JETP {\bf
  20}, 762 (1965)]}\BibitemShut {NoStop}%
\bibitem [{\citenamefont {Agterberg}\ \emph {et~al.}(2019)\citenamefont
  {Agterberg}, \citenamefont {Davis}, \citenamefont {Edkins}, \citenamefont
  {Fradkin}, \citenamefont {{Van Harlingen}}, \citenamefont {Kivelson},
  \citenamefont {Lee}, \citenamefont {Radzihovsky}, \citenamefont {Tranquada},\
  and\ \citenamefont {Wang}}]{Agterberg-2019}%
  \BibitemOpen
  \bibfield  {author} {\bibinfo {author} {\bibfnamefont {D.~F.}\ \bibnamefont
  {Agterberg}}, \bibinfo {author} {\bibfnamefont {J.~C.~S.}\ \bibnamefont
  {Davis}}, \bibinfo {author} {\bibfnamefont {S.~D.}\ \bibnamefont {Edkins}},
  \bibinfo {author} {\bibfnamefont {E.}~\bibnamefont {Fradkin}}, \bibinfo
  {author} {\bibfnamefont {D.~J.}\ \bibnamefont {{Van Harlingen}}}, \bibinfo
  {author} {\bibfnamefont {S.~A.}\ \bibnamefont {Kivelson}}, \bibinfo {author}
  {\bibfnamefont {P.~A.}\ \bibnamefont {Lee}}, \bibinfo {author} {\bibfnamefont
  {L.}~\bibnamefont {Radzihovsky}}, \bibinfo {author} {\bibfnamefont {J.~M.}\
  \bibnamefont {Tranquada}}, \ and\ \bibinfo {author} {\bibfnamefont
  {Y.}~\bibnamefont {Wang}},\ }\href@noop {} {\enquote {\bibinfo {title} {{The
  Physics of Pair Density Wave Superconductors: Cuprate Superconductors and
  Beyond}},}\ } (\bibinfo {year} {2019}),\ \bibinfo {note} {to appear in Annual
  Review of Condensed Matter Physics (2020)},\ \Eprint
  {http://arxiv.org/abs/arXiv:1904.09687} {arXiv:1904.09687} \BibitemShut
  {NoStop}%
\bibitem [{\citenamefont {Berg}\ \emph {et~al.}(2010)\citenamefont {Berg},
  \citenamefont {Fradkin},\ and\ \citenamefont {Kivelson}}]{berg2010}%
  \BibitemOpen
  \bibfield  {author} {\bibinfo {author} {\bibfnamefont {E.}~\bibnamefont
  {Berg}}, \bibinfo {author} {\bibfnamefont {E.}~\bibnamefont {Fradkin}}, \
  and\ \bibinfo {author} {\bibfnamefont {S.~A.}\ \bibnamefont {Kivelson}},\
  }\href {\doibase 10.1103/PhysRevLett.105.146403} {\bibfield  {journal}
  {\bibinfo  {journal} {Phys. Rev. Lett.}\ }\textbf {\bibinfo {volume} {105}},\
  \bibinfo {pages} {146403} (\bibinfo {year} {2010})}\BibitemShut {NoStop}%
\bibitem [{\citenamefont {Sikkema}\ \emph {et~al.}(1997)\citenamefont
  {Sikkema}, \citenamefont {Affleck},\ and\ \citenamefont
  {White}}]{Sikkema-1997}%
  \BibitemOpen
  \bibfield  {author} {\bibinfo {author} {\bibfnamefont {A.~E.}\ \bibnamefont
  {Sikkema}}, \bibinfo {author} {\bibfnamefont {I.}~\bibnamefont {Affleck}}, \
  and\ \bibinfo {author} {\bibfnamefont {S.~R.}\ \bibnamefont {White}},\ }\href
  {\doibase 10.1103/PhysRevLett.79.929} {\bibfield  {journal} {\bibinfo
  {journal} {Phys. Rev. Lett.}\ }\textbf {\bibinfo {volume} {79}},\ \bibinfo
  {pages} {929} (\bibinfo {year} {1997})}\BibitemShut {NoStop}%
\bibitem [{\citenamefont {Jaefari}\ and\ \citenamefont
  {Fradkin}(2012)}]{jaefari2012}%
  \BibitemOpen
  \bibfield  {author} {\bibinfo {author} {\bibfnamefont {A.}~\bibnamefont
  {Jaefari}}\ and\ \bibinfo {author} {\bibfnamefont {E.}~\bibnamefont
  {Fradkin}},\ }\href {\doibase 10.1103/PhysRevB.85.035104} {\bibfield
  {journal} {\bibinfo  {journal} {Phys. Rev. B}\ }\textbf {\bibinfo {volume}
  {85}},\ \bibinfo {pages} {035104} (\bibinfo {year} {2012})}\BibitemShut
  {NoStop}%
\bibitem [{\citenamefont {Zachar}\ and\ \citenamefont
  {Tsvelik}(2001)}]{Zachar-2001}%
  \BibitemOpen
  \bibfield  {author} {\bibinfo {author} {\bibfnamefont {O.}~\bibnamefont
  {Zachar}}\ and\ \bibinfo {author} {\bibfnamefont {A.~M.}\ \bibnamefont
  {Tsvelik}},\ }\href {\doibase 10.1103/PhysRevB.64.033103} {\bibfield
  {journal} {\bibinfo  {journal} {Phys. Rev. B}\ }\textbf {\bibinfo {volume}
  {64}},\ \bibinfo {pages} {033103} (\bibinfo {year} {2001})}\BibitemShut
  {NoStop}%
\bibitem [{\citenamefont {Zachar}(2001)}]{Zachar-2001b}%
  \BibitemOpen
  \bibfield  {author} {\bibinfo {author} {\bibfnamefont {O.}~\bibnamefont
  {Zachar}},\ }\href {\doibase 10.1103/PhysRevB.63.205104} {\bibfield
  {journal} {\bibinfo  {journal} {Phys. Rev. B}\ }\textbf {\bibinfo {volume}
  {63}},\ \bibinfo {pages} {205104} (\bibinfo {year} {2001})}\BibitemShut
  {NoStop}%
\bibitem [{\citenamefont {Cho}\ \emph {et~al.}(2014)\citenamefont {Cho},
  \citenamefont {Soto-Garrido},\ and\ \citenamefont {Fradkin}}]{cho2014}%
  \BibitemOpen
  \bibfield  {author} {\bibinfo {author} {\bibfnamefont {G.~Y.}\ \bibnamefont
  {Cho}}, \bibinfo {author} {\bibfnamefont {R.}~\bibnamefont {Soto-Garrido}}, \
  and\ \bibinfo {author} {\bibfnamefont {E.}~\bibnamefont {Fradkin}},\ }\href
  {\doibase 10.1103/PhysRevLett.113.256405} {\bibfield  {journal} {\bibinfo
  {journal} {Phys. Rev. Lett.}\ }\textbf {\bibinfo {volume} {113}},\ \bibinfo
  {pages} {256405} (\bibinfo {year} {2014})}\BibitemShut {NoStop}%
\bibitem [{\citenamefont {Read}\ and\ \citenamefont {Green}(2000)}]{Read-2000}%
  \BibitemOpen
  \bibfield  {author} {\bibinfo {author} {\bibfnamefont {N.}~\bibnamefont
  {Read}}\ and\ \bibinfo {author} {\bibfnamefont {D.}~\bibnamefont {Green}},\
  }\href {\doibase https://doi.org/10.1103/PhysRevB.61.10267} {\bibfield
  {journal} {\bibinfo  {journal} {Phys. Rev. B}\ }\textbf {\bibinfo {volume}
  {61}},\ \bibinfo {pages} {10267} (\bibinfo {year} {2000})}\BibitemShut
  {NoStop}%
\bibitem [{\citenamefont {Ivanov}(2001)}]{ivanov-2001}%
  \BibitemOpen
  \bibfield  {author} {\bibinfo {author} {\bibfnamefont {D.~A.}\ \bibnamefont
  {Ivanov}},\ }\href {\doibase https://doi.org/10.1103/PhysRevLett.86.268}
  {\bibfield  {journal} {\bibinfo  {journal} {Phys. Rev. Lett.}\ }\textbf
  {\bibinfo {volume} {86}},\ \bibinfo {pages} {268} (\bibinfo {year}
  {2001})}\BibitemShut {NoStop}%
\bibitem [{\citenamefont {Kitaev}(2001)}]{Kitaev-2001}%
  \BibitemOpen
  \bibfield  {author} {\bibinfo {author} {\bibfnamefont {A.~Y.}\ \bibnamefont
  {Kitaev}},\ }\href@noop {} {\bibfield  {journal} {\bibinfo  {journal}
  {Physics-Uspekhi}\ }\textbf {\bibinfo {volume} {44}},\ \bibinfo {pages} {131}
  (\bibinfo {year} {2001})},\ \bibinfo {note} {{Proceedings of the Mesoscopic
  and Strongly Correlated Electron Systems Conference (9-16 July 2000,
  Chernogolovka, Moscow, Russia)}}\BibitemShut {NoStop}%
\bibitem [{\citenamefont {Fu}\ and\ \citenamefont {Kane}(2008)}]{Fu-2008}%
  \BibitemOpen
  \bibfield  {author} {\bibinfo {author} {\bibfnamefont {L.}~\bibnamefont
  {Fu}}\ and\ \bibinfo {author} {\bibfnamefont {C.~L.}\ \bibnamefont {Kane}},\
  }\href {\doibase https://doi.org/10.1103/PhysRevLett.100.096407} {\bibfield
  {journal} {\bibinfo  {journal} {Phys. Rev. Lett.}\ }\textbf {\bibinfo
  {volume} {100}},\ \bibinfo {pages} {096407} (\bibinfo {year}
  {2008})}\BibitemShut {NoStop}%
\bibitem [{\citenamefont {Ruhman}\ \emph {et~al.}(2015)\citenamefont {Ruhman},
  \citenamefont {Berg},\ and\ \citenamefont {Altman}}]{Ruhman-2015}%
  \BibitemOpen
  \bibfield  {author} {\bibinfo {author} {\bibfnamefont {J.}~\bibnamefont
  {Ruhman}}, \bibinfo {author} {\bibfnamefont {E.}~\bibnamefont {Berg}}, \ and\
  \bibinfo {author} {\bibfnamefont {E.}~\bibnamefont {Altman}},\ }\href
  {\doibase 10.1103/PhysRevLett.114.100401} {\bibfield  {journal} {\bibinfo
  {journal} {Phys. Rev. Lett.}\ }\textbf {\bibinfo {volume} {114}},\ \bibinfo
  {pages} {100401} (\bibinfo {year} {2015})}\BibitemShut {NoStop}%
\bibitem [{\citenamefont {Tsvelik}(2016{\natexlab{a}})}]{tsvelik2016b}%
  \BibitemOpen
  \bibfield  {author} {\bibinfo {author} {\bibfnamefont {A.~M.}\ \bibnamefont
  {Tsvelik}},\ }\href@noop {} {\bibfield  {journal} {\bibinfo  {journal} {Phys.
  Rev. B}\ }\textbf {\bibinfo {volume} {94}},\ \bibinfo {pages} {205141}
  (\bibinfo {year} {2016}{\natexlab{a}})}\BibitemShut {NoStop}%
\bibitem [{\citenamefont {Witten}(1978)}]{Witten-1978}%
  \BibitemOpen
  \bibfield  {author} {\bibinfo {author} {\bibfnamefont {E.}~\bibnamefont
  {Witten}},\ }\href {\doibase https://doi.org/10.1016/0550-3213(78)90204-3}
  {\bibfield  {journal} {\bibinfo  {journal} {Nucl. Phys. B}\ }\textbf
  {\bibinfo {volume} {142}},\ \bibinfo {pages} {285} (\bibinfo {year}
  {1978})}\BibitemShut {NoStop}%
\bibitem [{\citenamefont {Shankar}(1985)}]{Shankar-1985}%
  \BibitemOpen
  \bibfield  {author} {\bibinfo {author} {\bibfnamefont {R.}~\bibnamefont
  {Shankar}},\ }\href {\doibase 10.1103/PhysRevLett.55.453} {\bibfield
  {journal} {\bibinfo  {journal} {Phys. Rev. Lett.}\ }\textbf {\bibinfo
  {volume} {55}},\ \bibinfo {pages} {453} (\bibinfo {year} {1985})}\BibitemShut
  {NoStop}%
\bibitem [{\citenamefont {Witten}(1984)}]{Witten-1984}%
  \BibitemOpen
  \bibfield  {author} {\bibinfo {author} {\bibfnamefont {E.}~\bibnamefont
  {Witten}},\ }\href {\doibase 10.1007/BF01215276} {\bibfield  {journal}
  {\bibinfo  {journal} {Comm. Math. Phys.}\ }\textbf {\bibinfo {volume} {92}},\
  \bibinfo {pages} {455} (\bibinfo {year} {1984})}\BibitemShut {NoStop}%
\bibitem [{\citenamefont {Balents}\ and\ \citenamefont
  {Fisher}(1996)}]{Balents-1996}%
  \BibitemOpen
  \bibfield  {author} {\bibinfo {author} {\bibfnamefont {L.}~\bibnamefont
  {Balents}}\ and\ \bibinfo {author} {\bibfnamefont {M.~P.~A.}\ \bibnamefont
  {Fisher}},\ }\href {\doibase 10.1103/PhysRevB.53.12133} {\bibfield  {journal}
  {\bibinfo  {journal} {Phys. Rev. B}\ }\textbf {\bibinfo {volume} {53}},\
  \bibinfo {pages} {12133} (\bibinfo {year} {1996})}\BibitemShut {NoStop}%
\bibitem [{\citenamefont {Schollw{\"{o}}ck}(2011)}]{schollwock2011}%
  \BibitemOpen
  \bibfield  {author} {\bibinfo {author} {\bibfnamefont {U.}~\bibnamefont
  {Schollw{\"{o}}ck}},\ }\href {\doibase 10.1016/j.aop.2010.09.012} {\bibfield
  {journal} {\bibinfo  {journal} {Ann. Phys. (N. Y).}\ }\textbf {\bibinfo
  {volume} {326}},\ \bibinfo {pages} {96} (\bibinfo {year} {2011})}\BibitemShut
  {NoStop}%
\bibitem [{\citenamefont {Pollmann}\ \emph {et~al.}(2010)\citenamefont
  {Pollmann}, \citenamefont {Turner}, \citenamefont {Berg},\ and\ \citenamefont
  {Oshikawa}}]{Pollmann-2010}%
  \BibitemOpen
  \bibfield  {author} {\bibinfo {author} {\bibfnamefont {F.}~\bibnamefont
  {Pollmann}}, \bibinfo {author} {\bibfnamefont {A.~M.}\ \bibnamefont
  {Turner}}, \bibinfo {author} {\bibfnamefont {E.}~\bibnamefont {Berg}}, \ and\
  \bibinfo {author} {\bibfnamefont {M.}~\bibnamefont {Oshikawa}},\ }\href
  {\doibase 10.1103/PhysRevB.81.064439} {\bibfield  {journal} {\bibinfo
  {journal} {Phys. Rev. B}\ }\textbf {\bibinfo {volume} {81}},\ \bibinfo
  {pages} {064439} (\bibinfo {year} {2010})}\BibitemShut {NoStop}%
\bibitem [{\citenamefont {Robinson}\ \emph {et~al.}(2019)\citenamefont
  {Robinson}, \citenamefont {Altland}, \citenamefont {Egger}, \citenamefont
  {Gergs}, \citenamefont {Li}, \citenamefont {Schuricht}, \citenamefont
  {Tsvelik}, \citenamefont {Weichselbaum},\ and\ \citenamefont
  {Konik}}]{robinson2019}%
  \BibitemOpen
  \bibfield  {author} {\bibinfo {author} {\bibfnamefont {N.~J.}\ \bibnamefont
  {Robinson}}, \bibinfo {author} {\bibfnamefont {A.}~\bibnamefont {Altland}},
  \bibinfo {author} {\bibfnamefont {R.}~\bibnamefont {Egger}}, \bibinfo
  {author} {\bibfnamefont {N.~M.}\ \bibnamefont {Gergs}}, \bibinfo {author}
  {\bibfnamefont {W.}~\bibnamefont {Li}}, \bibinfo {author} {\bibfnamefont
  {D.}~\bibnamefont {Schuricht}}, \bibinfo {author} {\bibfnamefont {A.~M.}\
  \bibnamefont {Tsvelik}}, \bibinfo {author} {\bibfnamefont {A.}~\bibnamefont
  {Weichselbaum}}, \ and\ \bibinfo {author} {\bibfnamefont {R.~M.}\
  \bibnamefont {Konik}},\ }\href {\doibase 10.1103/PhysRevLett.122.027201}
  {\bibfield  {journal} {\bibinfo  {journal} {Phys. Rev. Lett.}\ }\textbf
  {\bibinfo {volume} {122}},\ \bibinfo {pages} {027201} (\bibinfo {year}
  {2019})}\BibitemShut {NoStop}%
\bibitem [{\citenamefont {Tsvelik}(2016{\natexlab{b}})}]{tsvelik2016}%
  \BibitemOpen
  \bibfield  {author} {\bibinfo {author} {\bibfnamefont {A.~M.}\ \bibnamefont
  {Tsvelik}},\ }\href {\doibase https://doi.org/10.1103/PhysRevB.94.165114}
  {\bibfield  {journal} {\bibinfo  {journal} {Phys. Rev. B}\ }\textbf {\bibinfo
  {volume} {94}},\ \bibinfo {pages} {165114} (\bibinfo {year}
  {2016}{\natexlab{b}})}\BibitemShut {NoStop}%
\bibitem [{\citenamefont {{Di Francesco}}\ \emph {et~al.}(1997)\citenamefont
  {{Di Francesco}}, \citenamefont {Mathieu},\ and\ \citenamefont
  {S\'en\'echal}}]{Difrancesco1997}%
  \BibitemOpen
  \bibfield  {author} {\bibinfo {author} {\bibfnamefont {P.}~\bibnamefont {{Di
  Francesco}}}, \bibinfo {author} {\bibfnamefont {P.}~\bibnamefont {Mathieu}},
  \ and\ \bibinfo {author} {\bibfnamefont {D.}~\bibnamefont {S\'en\'echal}},\
  }\href@noop {} {\emph {\bibinfo {title} {{Conformal Field Theory}}}}\
  (\bibinfo  {publisher} {Springer-Verlag},\ \bibinfo {address} {Berlin},\
  \bibinfo {year} {1997})\BibitemShut {NoStop}%
\bibitem [{\citenamefont {Polchinski}(1998)}]{Polchinski-1998}%
  \BibitemOpen
  \bibfield  {author} {\bibinfo {author} {\bibfnamefont {J.}~\bibnamefont
  {Polchinski}},\ }\href@noop {} {\emph {\bibinfo {title} {{String Theory}}}},\
  \bibinfo {edition} {1st}\ ed.,\ Vol.~\bibinfo {volume} {II}\ (\bibinfo
  {publisher} {Cambridge University Press},\ \bibinfo {address} {Cambridge,
  UK},\ \bibinfo {year} {1998})\BibitemShut {NoStop}%
\bibitem [{\citenamefont {Affleck}\ and\ \citenamefont
  {Haldane}(1987)}]{Affleck-1987}%
  \BibitemOpen
  \bibfield  {author} {\bibinfo {author} {\bibfnamefont {I.}~\bibnamefont
  {Affleck}}\ and\ \bibinfo {author} {\bibfnamefont {F.~D.~M.}\ \bibnamefont
  {Haldane}},\ }\href {\doibase https://doi.org/10.1103/PhysRevB.36.5291}
  {\bibfield  {journal} {\bibinfo  {journal} {Phys. Rev. B}\ }\textbf {\bibinfo
  {volume} {36}},\ \bibinfo {pages} {5291} (\bibinfo {year}
  {1987})}\BibitemShut {NoStop}%
\bibitem [{\citenamefont {Shelton}\ \emph {et~al.}(1996)\citenamefont
  {Shelton}, \citenamefont {Nersesyan},\ and\ \citenamefont
  {Tsvelik}}]{shelton1996}%
  \BibitemOpen
  \bibfield  {author} {\bibinfo {author} {\bibfnamefont {D.~G.}\ \bibnamefont
  {Shelton}}, \bibinfo {author} {\bibfnamefont {A.~A.}\ \bibnamefont
  {Nersesyan}}, \ and\ \bibinfo {author} {\bibfnamefont {A.~M.}\ \bibnamefont
  {Tsvelik}},\ }\href {\doibase https://doi.org/10.1103/PhysRevB.53.8521}
  {\bibfield  {journal} {\bibinfo  {journal} {Phys. Rev. B}\ }\textbf {\bibinfo
  {volume} {53}},\ \bibinfo {pages} {8521} (\bibinfo {year}
  {1996})}\BibitemShut {NoStop}%
\bibitem [{\citenamefont {Allen}\ and\ \citenamefont
  {S{\'e}n{\'e}chal}(1997)}]{allen1997}%
  \BibitemOpen
  \bibfield  {author} {\bibinfo {author} {\bibfnamefont {D.}~\bibnamefont
  {Allen}}\ and\ \bibinfo {author} {\bibfnamefont {D.}~\bibnamefont
  {S{\'e}n{\'e}chal}},\ }\href {\doibase
  https://doi.org/10.1103/PhysRevB.55.299} {\bibfield  {journal} {\bibinfo
  {journal} {Phys. Rev. B}\ }\textbf {\bibinfo {volume} {55}},\ \bibinfo
  {pages} {299} (\bibinfo {year} {1997})}\BibitemShut {NoStop}%
\bibitem [{\citenamefont {Lecheminant}\ and\ \citenamefont
  {Orignac}(2002)}]{lecheminant2002}%
  \BibitemOpen
  \bibfield  {author} {\bibinfo {author} {\bibfnamefont {P.}~\bibnamefont
  {Lecheminant}}\ and\ \bibinfo {author} {\bibfnamefont {E.}~\bibnamefont
  {Orignac}},\ }\href {\doibase https://doi.org/10.1103/PhysRevB.65.174406}
  {\bibfield  {journal} {\bibinfo  {journal} {Phys. Rev. B}\ }\textbf {\bibinfo
  {volume} {65}},\ \bibinfo {pages} {174406} (\bibinfo {year}
  {2002})}\BibitemShut {NoStop}%
\bibitem [{\citenamefont {Jackiw}\ and\ \citenamefont
  {Rebbi}(1976)}]{Jackiw-1976}%
  \BibitemOpen
  \bibfield  {author} {\bibinfo {author} {\bibfnamefont {R.}~\bibnamefont
  {Jackiw}}\ and\ \bibinfo {author} {\bibfnamefont {C.}~\bibnamefont {Rebbi}},\
  }\href {\doibase https://doi.org/10.1103/PhysRevD.13.3398} {\bibfield
  {journal} {\bibinfo  {journal} {Phys. Rev. D}\ }\textbf {\bibinfo {volume}
  {13}},\ \bibinfo {pages} {3398} (\bibinfo {year} {1976})}\BibitemShut
  {NoStop}%
\bibitem [{\citenamefont {Santos}\ \emph {et~al.}(2019)\citenamefont {Santos},
  \citenamefont {Wang},\ and\ \citenamefont {Fradkin}}]{santos2019pair}%
  \BibitemOpen
  \bibfield  {author} {\bibinfo {author} {\bibfnamefont {L.~H.}\ \bibnamefont
  {Santos}}, \bibinfo {author} {\bibfnamefont {Y.}~\bibnamefont {Wang}}, \ and\
  \bibinfo {author} {\bibfnamefont {E.}~\bibnamefont {Fradkin}},\ }\href
  {\doibase 10.1103/PhysRevX.9.021047} {\bibfield  {journal} {\bibinfo
  {journal} {Phys. Rev. X}\ }\textbf {\bibinfo {volume} {9}},\ \bibinfo {pages}
  {021047} (\bibinfo {year} {2019})}\BibitemShut {NoStop}%
\bibitem [{ite()}]{itensor}%
  \BibitemOpen
  \href@noop {} {}\bibinfo {howpublished}
  {\url{http://itensor.org/}}\BibitemShut {NoStop}%
\bibitem [{\citenamefont {Affleck}(1990)}]{Affleck-1988}%
  \BibitemOpen
  \bibfield  {author} {\bibinfo {author} {\bibfnamefont {I.}~\bibnamefont
  {Affleck}},\ }in\ \href@noop {} {\emph {\bibinfo {booktitle} {{Strings,
  Fields and Critical Phenomena}}}},\ \bibinfo {series and number} {Proceedings
  of the Les Houches Summer School 1988, {Session XLIX}, E. Br\'ezin and J.
  Zinn-Justin editors}\ (\bibinfo  {publisher} {North-Holland},\ \bibinfo
  {address} {Amsterdam, the Netherlands},\ \bibinfo {year} {1990})\ p.\
  \bibinfo {pages} {563}\BibitemShut {NoStop}%
\bibitem [{\citenamefont {Wu}\ \emph {et~al.}(2003)\citenamefont {Wu},
  \citenamefont {Liu},\ and\ \citenamefont {Fradkin}}]{wu2003}%
  \BibitemOpen
  \bibfield  {author} {\bibinfo {author} {\bibfnamefont {C.}~\bibnamefont
  {Wu}}, \bibinfo {author} {\bibfnamefont {W.~V.}\ \bibnamefont {Liu}}, \ and\
  \bibinfo {author} {\bibfnamefont {E.}~\bibnamefont {Fradkin}},\ }\href
  {\doibase https://doi.org/10.1103/PhysRevB.68.115104} {\bibfield  {journal}
  {\bibinfo  {journal} {Phys. Rev. B}\ }\textbf {\bibinfo {volume} {68}},\
  \bibinfo {pages} {115104} (\bibinfo {year} {2003})}\BibitemShut {NoStop}%
\end{thebibliography}%
\bibliographystyle{apsrev4-1}

\end{document}